\documentclass[12pt,preprint]{aastex}

\begin{document}

\title{Star Formation via the Little Guy:\\
A Bayesian Study of Ultracool Dwarf Imaging Surveys for Companions}

\author{Peter R.\ Allen}

\affil{Pennsylvania State University, Dept.\ of Astronomy and Astrophysics, 525 Davey Lab, University Park, PA 16802; pallen@astro.psu.edu}

\begin{abstract}

I have undertaken a comprehensive statistical investigation of the ultracool dwarf companion distribution (spectral type M6 and later).  Utilizing a Bayesian algorithm, I tested models of the companion distribution against data from an extensive set of space and ground-based imaging observations of nearby ultracool dwarfs.  My main conclusions are fivefold: 1) Confirm that the concentration of high mass ratio ultracool binary systems is a fundamental feature of the companion distribution, not an observational or selection bias; 2) Determine that the wide ($>{\sim}20$~AU) binary frequency can be no more the $1-2\%$; 3) Show that the decreasing binary frequency with later spectral types is a real trend; 4) Demonstrate that a large population of currently undetected low mass ratio systems are not consistent with the current data; 5) Find that the population of spectroscopic binaries must be be at least 30\% that of currently known ultracool binaries.  The best fit value for the overall M6 and later binary frequency is ${\sim}20\%-22\%$, of which only $\sim$6\% consists of currently undetected companions with separations less than 1~AU.  If this is correct, then the upper limit of the ultracool binary population discovered to date is $\sim$75\%.  I find that the numerical simulation results of the ejection formation method are inconsistent with the outcome of this analysis.  However, dynamics do seem to play an important role as simulations of small-N clusters and triple system decays produce results similar to those of this work.  The observational efforts required to improve these constraints are shown to be primarily large spectroscopic binary surveys and improved high-resolution imaging techniques.

\end{abstract}


\section{Introduction}

An understanding of the star formation process has long been a major goal of modern astronomy.  The pioneering work of \citet{salp} established the Initial Mass Function (IMF), the number of stars born per unit mass per unit volume, as a fundamental distribution for star formation theory.  The study of the IMF was later expanded by the work of \citet{ms79} and countless efforts since in many young clusters and the field.  A number of functional forms of the IMF have been tested in varied environments, such as the lognormal \citep{ms79} and the power-law \citep{salp}.  As a community, we are convinced that there is a link between the IMF and the global process by which those stars formed.  However, there is another set of by-products of star formation that can elucidate the formation process: binary companions.

The frequency and distribution of physical properties of companions to intermediate to low-mass stars has been well characterized.  \citet{dm} showed that G stars have a binary rate of ${\sim}60\%$ with a Gaussian distribution of separations.  \citet{fm} performed similar studies but for early-M dwarfs and found a binary rate of ${\sim}40\%$ with a similar Gaussian distribution in separation to the G dwarf studies.  Both of these studies used samples drawn from catalogs of known objects that have their own selection biases.  They also used multiple observational techniques to derive their binary frequencies and companion distributions.  As a result, the frequencies derived incorporate corrections to account for the inherent uncertainties in the original samples and their combination.  This is an unfortunate, but necessary step in most companion analyzes (including this one).  Hotter stars binary frequencies are not as well understood.  This is mostly because there are relatively few very massive stars and they evolve very quickly.  However, there is a well-established trend that the binary frequency increase with mass at the high end.  For example, the binary frequency of O stars is at least 75\% and most likely much higher \citep{bm98}.

Recent years have ushered in a new area of binary studies: very low-mass stars ($<0.15M_{\odot}$), which includes ultracool dwarfs (spectral type M7 and later), many of the coolest and lowest mass extending into the brown dwarf regime.  After years of fruitless searching for brown dwarfs, the profusion of deep far-red optical (Sloan Digital Sky Survey, SDSS \citet{york}) and near-infrared (Deep Near Infrared Survey, DENIS \citet{denis}, and the Two-Micron All Sky Survey, 2MASS \citet{tmass}) sky surveys in the last decade catalyzed the discovery of hundreds of nearby ultracool dwarfs. This enables us to probe an entirely new mass range of the star formation process.  

Shortly after the discovery of this large population of nearby ultracool dwarfs in the field, the community began to search for companions to them, see \citet{burg07} or \citet{duch07} for a review.  The observed field ultracool dwarf companion distribution has been shown to be quite different from that of higher mass stars.  These imaging surveys have shown that $\sim$15\% of late-M and L dwarfs are resolved as binary systems.  Only a handful of these systems has a separation exceeding 15~AU.  This is in contrast to ultracool dwarfs with higher mass primaries; for example, VB~10, the archetypal late-type M dwarf, lies 400~AU from its primary, the M3 dwarf, Gl~752A \citep{vb}, while the nearby T dwarfs, Gl~229B \citep{naka}, Gl~570D \citep{burg00}, and $\epsilon$ Indi Bab \citep{eindi}, are all wide components in multiple systems.  On the other hand, a handful of wide ($>50$~AU), predominantly young ($<$10~Myr), ultracool binaries have been identified.  \citet{lu04} has identified a pair of late-type M dwarfs separated by 240~AU in $\rho$ Ophiuchus, an additional 212~AU M dwarf pair in Ophiuchus was found by \citet{close06}, \citet{chau04} have discovered a very low-mass brown dwarf companion of the TW~Hya member 2M1207$-$39, with a separation of 60~AU, and \citet{luh06} has found a 210~AU pairing in Chameleon I.  There are two field binary systems discovered so far: DENIS 0551$-$4434AB a M8.5/L0 pair with a separation of over 200~AU \citep{bill05}; and 2MASS 0126AB, a M6.5/M8 pair with a separation of 5100~AU \citep{art07}.  It should be noted that \citet{art07} argue that 2MASS 0126AB may be part of the TW Hydra association, and thus may be a young, wide binary like 2MASS 1207-39.

\begin{figure}
\centering
\includegraphics[scale=0.7,angle=-90]{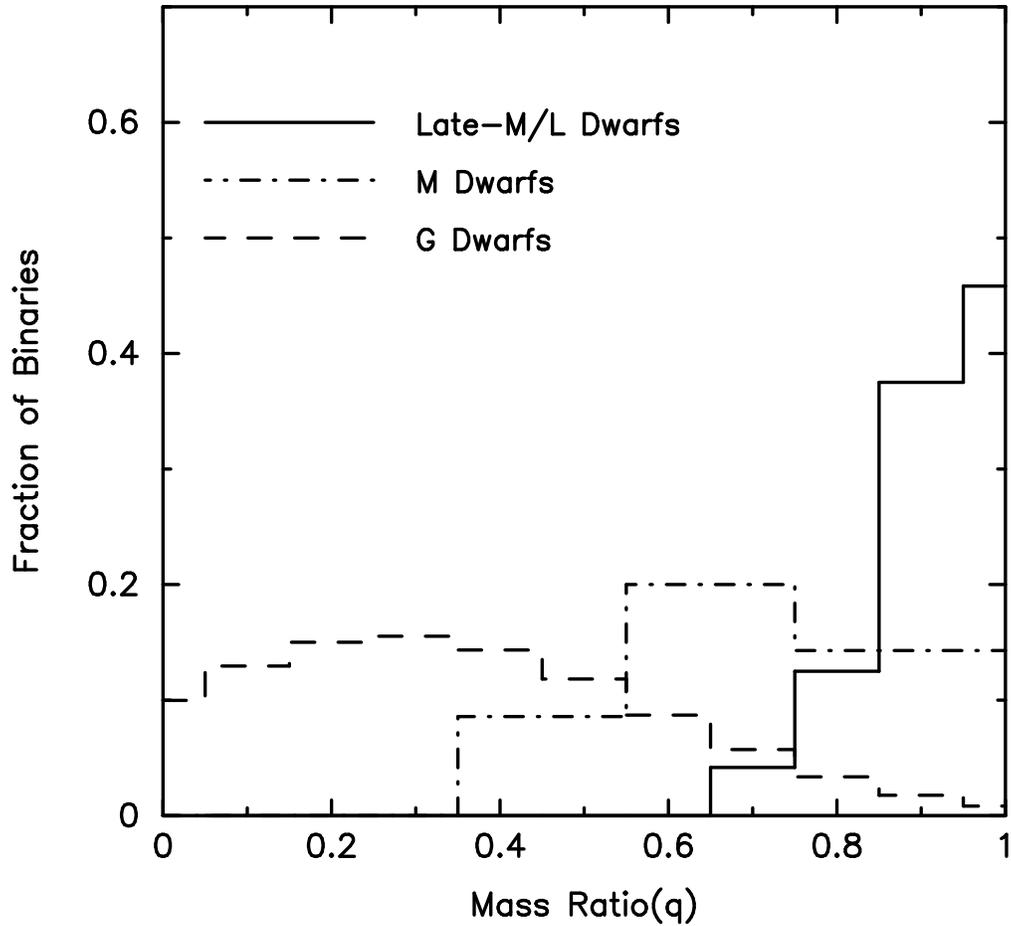}
\caption{Comparison of mass ratio distributions for G stars (\citet{dm}, dashed), early-M stars (\citet{fm}, dot-dashed), and currently known ultracool systems (\citet{burg07}, solid).  Note that the mass ratio distribution appears to evolve from higher to lower mass primaries.}
\label{fig:km2}
\end{figure}

\begin{figure}
\centering
\includegraphics[scale=0.7,angle=-90]{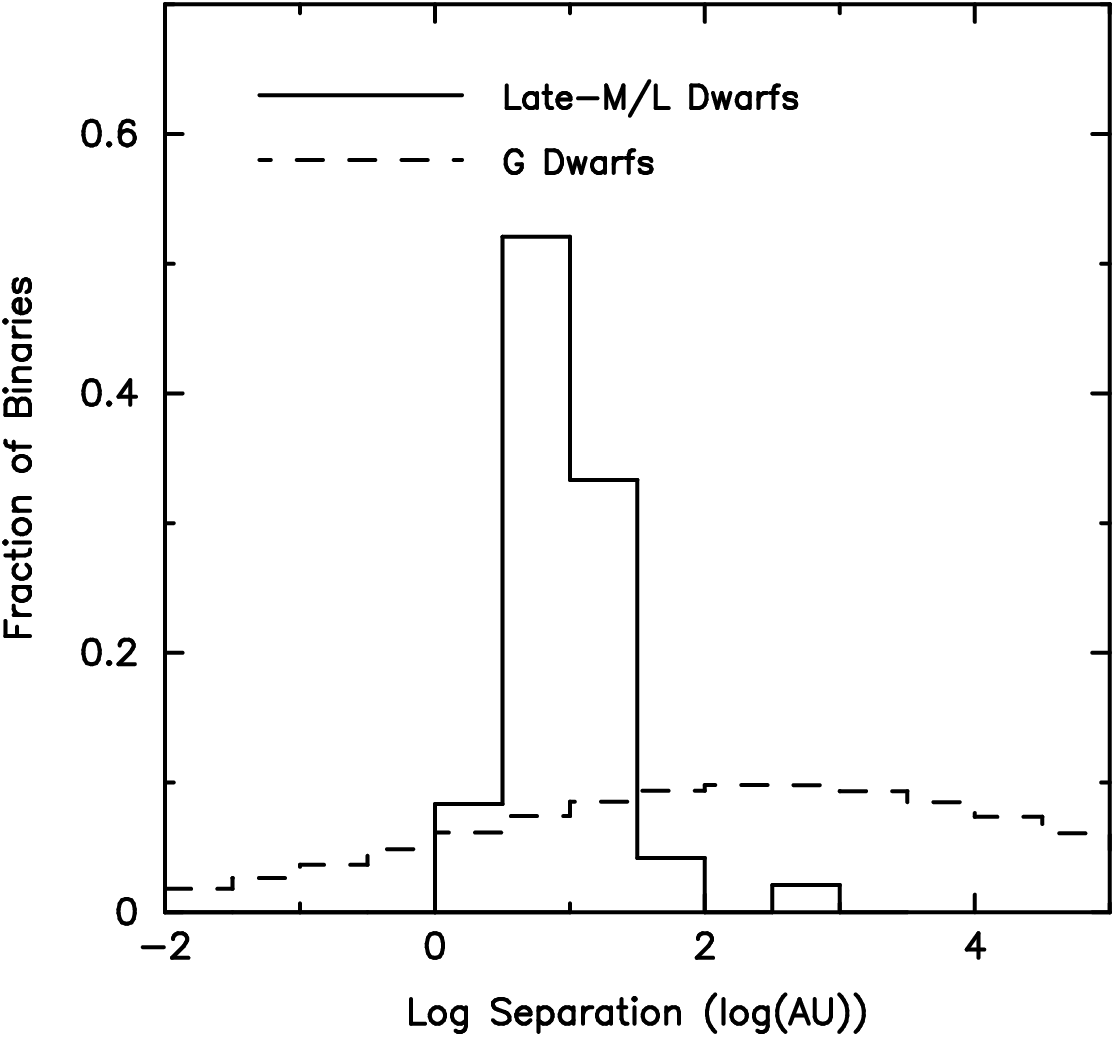}
\caption{Histograms of the separation distribution of all known companions to late-M and L field dwarfs (\citet{burg07}, solid), and for comparison companions to G-dwarfs (\citet{dm}, dashed).  Note that the separation distribution of early-M dwarfs is very similar to that of G dwarfs, according to \citet{fm} and is not plotted separately here.  This plot demonstrates the lack of wide companions as compared to the stellar companion population and the Gaussian shape of the ultracool dwarf separation distribution.}
\label{fig:kc}
\end{figure}

The detected field binary systems are predominantly equal brightness/mass systems (Figure \ref{fig:km2}) in tightly bound orbits (Figure \ref{fig:kc}) with relatively few exceptions.  However, we would like to know if these features only due to observational/selection biases or are real features of the ultracool dwarf companion distribution.  For example, the majority of the imaging surveys from which these conclusions are drawn are based on magnitude limited samples.  Such samples tend to pick out the brightest objects of a given spectral class.  Equal brightness/mass binaries are a prime contaminant because they appear to be very bright when the are in fact two separate objects.  In any analysis of the overall distribution of ultracool dwarf binaries such biases must be taken into account.  In this work, I will use techniques based on those derived in \citet{burg03}.  

Numerical simulations of the various formation scenarios are just beginning to produce predictions of the binary properties of samples formed via those mechanisms, see \citet{burg07} for a review of formation mechanisms and simulation results.  These simulations encompass a wide variety, including standard cloud fragmentation, massive circumstellar disk instabilities, and photo-evaporation by nearby OB stars.  I will compare the outcome of my analysis to the few simulations that produce testable results \citep{bbb02,bbb03,bb05,ster,um}.

To date there has not yet been a comprehensive statistical analysis of the ultracool dwarf companion distribution.  This work will provide the first such study for companions to nearby field late-M, L, and T dwarfs, using only imaging data.  I use a Bayesian algorithm to determine the underlying companion distribution, utilizing both the detections in the surveys as well as the non-detections.  The Bayesian method also allows the easy incorporation of multiple data sets with one another to generate a unified conclusion.  

In this paper I will bring together the disparate ultracool binary imaging data sets with theoretical models that describe the evolution of brown dwarfs \citep{bur} to constrain the ultracool binary distribution via a Bayesian statistical approach similar to that found in \citet{a05}, hereafter referred to as A05.  I will also determine if the unusual features in the ultracool binary distribution (the paucity of wide systems and the preference for near-equal masses) are real or are an observational/selection bias.  This paper is organized as follows: Section 2 discusses the binary surveys to be used in my statistical analysis. Section 3 explains my basic Bayesian methodology and Section 4 outlines how I produce the binary distribution models.  Section 5 discusses the results and Section 6 summarizes my conclusions.

\section{Summary of Companion Surveys and Data Sets}
\label{sec:survey}

\subsection{Surveys and Detections}

As will be seen in Section \ref{sec:ba}, the Bayesian algorithm requires data sets that report a complete list of discovered binaries as well as non-detections.  This eliminates systems discovered serendipitously from this analysis.  Additionally, I will focus this work on present imaging results, thus excluding spectroscopic binaries.  The reasons for this are twofold: 1) The spectroscopic surveys are not yet as fully developed as the imaging surveys in size and scope; 2) A great deal of additional modeling is needed to properly include the spectroscopic results and it was deemed outside the reach of this paper.  The analysis of spectroscopic results will be the topic of a future paper.  Two examples of ultracool binaries to be excluded are PPl 15AB \citep{basri} and $\epsilon$IndiBab, the former was discovered via high resolution spectral measurements of a brown dwarf candidate in the Pleiades and the latter serendipitously found in a proper motion survey.  The serendipitous binaries cannot be included because the Bayesian algorithm I use requires a full set of detections and non-detections.  Such data cannot be obtained from a chance discovery.  In all $\sim$20\% of known ultracool dwarf binary systems fall into the excluded category.  This is based on the compiled list of 75 ultracool binaries in Table 1 of \citet{burg07} and eliminating those found either serendipitously or in non-imaging surveys.  

The majority of the ultracool dwarf companion surveys have been carried out with HST.  The first set, using WFPC2, is summarized in \citet{bouy03}, and is the combined work of \citet{inr01}, \citet{gizis03}, and \citet{bouy03}.  These groups surveyed a total of 131 late-M and L dwarfs with HST and found 25 binary systems.  \citet{burg03} also used WFPC2 to examine 10 T dwarfs and resolved 2 binaries.  Additional HST surveys, using the NICMOS NIC1 camera, obtained high resolution images of 3 M dwarfs and 49 L dwarfs that are mostly within 20~pc of the Sun \citep{inr06}, and 22 T dwarfs by \citet{burg06m}.  These observations resolved a further 14 binaries.  The other high-resolution surveys \citep{close03,ns03,ns05} are a ground-based program that used the near-IR adaptive optics system at Gemini-North, Hokupa'a \citep{hokupaa}.  \citet{close03} examined 39 M8 through L0 dwarfs and found 9 binary systems, and \citet{ns03,ns05} examined 41 M6-M7.5 dwarfs and resolved 5 systems.

This analysis also uses data from two low-resolution, wide-field, ground-based surveys of late-M and L dwarfs for companions.  The preliminary results of a survey using the Keck Telescope \citep{k99}, with the Near InfraRed Camera (NIRC) \citep{nirc}, found 3 binary systems out of ten surveyed.  The full data set ($76$ objects) is not yet published (Koerner et~al., in prep.) and includes one other binary system, which was previously resolved by \citet{inr01}.  All four of these binary systems have also been observed within the HST programs described above.  The final survey is one that used NSFCam on NASA's IRTF to study 133 late-M and L dwarfs and found no candidate companions \citep{a07}.

In all, a total 361 individual targets were imaged with these surveys, including 53 that are resolved as binary systems, which yields an observed binary fraction of $\sim$15\%.  Figure \ref{fig:spt} displays the spectral type and distance distributions of the primary star population for the above surveys.  These systems represent a large cross-section of the late-M., L, and T dwarf populations at the time of each survey, all with selection biases.  The early surveys \citep{k99,inr01,gizis03,bouy03,burg03,burg06m,a07} observed all the objects they knew about at the time regardless of distance or selection effects.  The AO surveys of \citet{close03,ns03,ns05} were constrained to the brightest and closest objects based on the Hokupa'a natural guide star AO system they were using.  The most recently devised and implemented survey of \citet{inr06} has a much better defined sample of ultracool dwarfs.  These objects were primarily taken from those found in \citet{klc03}, and represents a volume-limited sample of early and mid L dwarfs within 20~pc of the Sun.

The binaries found thus far point to a companion distribution that differs from that of higher mass primaries, G, K, and early-M stars.  Figure \ref{fig:km2} displays a comparison of the mass ratio distribution of the detected binaries in several nearby companion populations.  The trend in these data, continued by the ultracool dwarf companions, is to an increasingly peaked mass ratio distribution with decreasing primary mass.  It remains possible, though unlikely, that this is due the observational bias towards equal brightness binaries.  Recent work of \citet{burg06m} demonstrates that, when this bias is taken into account, the observed ultracool binary frequency is considerably lower than that of higher mass primaries when compared in similar regions of parameter space.  Thus, this trend appears to be genuine.

\subsection{Window Functions and Bias Correction}

The surveys included here are not sensitive to low-mass ratio companions at small separations and they suffer from decreasing linear resolution with increasing primary distance.  This can be seen in Figure \ref{fig:winfunc}, which displays the observed window functions of all the surveys used in this analysis.  I define the window function as the area in observational space (delta magnitude vs.\ the log of the projected linear separation) in which a companion could have been detected.  This was determined for each individual target based on information in their respective source papers.  For the HST WFPC2 surveys of late-M and L dwarfs I used the typical sensitivity limits given in \citet{gizis03}, and the WFPC2 surveys of T dwarfs used the individual sensitivity estimates given in \citet{burg03}.  The HST NICMOS L dwarf survey \citep{inr06} provided a sensitivity curve as a function of radial distance from the central source and was used to prepare that window function.  The NICMOS T dwarf survey \citep{burg07} again provided individual sensitivity measurements for each target.  For the AO surveys \citep{close03,ns03,ns05}, I used the sensitivity curve provided in \citet{ns05} for all the observations, as the author indicated that the curve was representative (Nick Siegler, private communication).  The Keck survey also contains individual sensitivity measurements for each target and they were used in the construction of their window functions (David Koerner, private communication).  Finally, the IRTF survey \citep{a07} also provides individual sensitivity measurements.

The window functions displayed in Figure \ref{fig:winfunc} are constructed as follows.  In each bin of observable space (delta mag vs.\ projected separation) I examine the sensitivity curves described above for the appropriate survey to which a target belongs.  If that bin is observable, it is assigned a 1, if not it is assigned a 0.  The only exception to this assignment is near the edges of each field.  All of these surveys were conducted with square fields of view (FOV).  However, the projected separation is not probed uniformly out to the value represented at the corner of the field.  Consequently, I created a weighting function that assumes a circular FOV out to a radius equal to half the diagonal of the square FOV which scales the area probed outside of a radius equal to the FOV/2.  This yields a gradual drop off of sensitivity to larger projected separations.  The final modification of the window function is to correct for the selection bias of the samples from which the targets were compiled.  Recall, that magnitude limited surveys are biased towards the brightest objects, which are more likely to be unresolved binary systems.  To correct for this, I use a technique described in \citet{burg03}.  They derive the following form to correct this bias as a function of delta magnitude:
\begin{equation}
(1 + 10^{0.4{\times}{\Delta}m})^{1.5}
\label{eq:selbias}
\end{equation}
The result of this correction is to decrease the significance of detections at small delta magnitudes, or near-equal mass systems, as required.

The general structure in the window functions is dominated by a step-like shape at small separations.  This is the result of decreasing sensitivity to low mass ratio binaries the closer you get to the central source.  The cutoff at a delta magnitude of 5 in the HST WFPC2 window function is due to the average sensitivity curve used for each observation in that survey.  This is different from the K-band window function which is a composite of four different surveys \citep{close03,ns03,ns05,a07,k99}.  These surveys were able to probe to smaller mass ratios (larger differences between primary and secondary mass) because smaller companions are brighter in the near-IR photometry bands than in the optical WFPC2 bands.  Also, both Koerner et~al.\ (in prep) and \citet{a07} provide individual sensitivity measurements for each target which leads to a non-uniform cutoff at large delta magnitudes.  Overall, there are two major holes in the coverage of these window functions: 1) very wide, low mass ratio companions and 2) sub-AU separation companions.  As a result, there may be a population of tight or wide, low-mass companions hidden where they cannot be seen in the current surveys.  These potential populations are discussed in Section \ref{sec:res}.

\section{Models of the Substellar Companion Distribution}
\label{sec:cmod}

The first step in the Bayesian statistical analysis is to build a physical companion distribution model that I will test against the data.  There are two main components to consider in a companion distribution, how they are distributed around each star (orbital parameters) and their relative masses (mass ratio).  Here I setup and justify those model distributions.

\subsection{Separation Distribution}

The separation distribution of ultracool dwarf binaries will be modeled as a Gaussian.  This is based on existing data for companions to GKM stars, as described in Sections 1 and 2.  The projected separation distribution of companions to GKM stars were found to be consistent with a log-normal (see Figure 7 of \citet{dm}, Figure 1b of \citet{mayor}, and Figure 2b of \citet{fm}).  Figure \ref{fig:kc} displays the separations of ultracool binaries, as summarized in \citet{burg07}, superimposed on the G star binaries of \citet{dm}.  While the small separations are incomplete, the overall observed distribution has been shown in the work of \citet{burg07} to be fit well by a symmetric Gaussian.  This is similar to the separation distributions derived for higher mass primaries \citep{dm,fm}.  However, I will also consider an asymmetric Gaussian shape, as described in \citet{jm05}.  They use this shape to statistically examine the likelihood of spectroscopic binaries at separations less than 1-2 AU and introduce a sharp cutoff at 10-20 AU to account for the lack of wide companions.  In interest of completeness, I will test this model as well.

The parameterized model form of the standard symmetric Gaussian is straightforward:
\begin{equation}
P(\log(a)|\log(a_0),{\sigma}_a){\Delta}{\log}(a) = \frac{1}{\sqrt{2\pi}{\sigma}_a}e^{-\frac{(\log(a) - \log(a_0))^2}{2{\sigma_a}^2}}{\Delta}{\log}(a)
\label{eq:sepinit}
\end{equation}
where $\log(a)$ is the log of the semi-major axis (AU) of the orbit, ${\Delta}\log(a)$ is the bin size in separation, $\log(a_0)$ defines the center point of the log-normal separation model, and $\sigma_a$ is the width.  The asymmetric Gaussian model of \citet{jm05} is constructed by multiplying Equation \ref{eq:sepinit} by a strong exponential cutoff beginning at 10~AU.
\begin{equation}
P(\log(a)|\log(a_0),{\sigma}_a){\Delta}{\log}(a) = \frac{1}{\sqrt{2\pi}{\sigma}_a}e^{-\frac{(\log(a) - \log(a_0))^2}{2{\sigma_a}^2}}e^{(1-\log(a))^3}{\Delta}{\log}(a)
\label{eq:sepinit2}
\end{equation}

However, the semi-major axis is not what is measured in a binary system; it is the projected separation on the plane of the sky.  In order to transform the above model, the companions are randomly distributed in inclination, phase, and eccentricity.  The projection onto the plane of the sky for a particular value of the inclination angle, $i$, the phase angle, $\phi$, and the eccentricity, $e$, is given by the following form:
\begin{equation}
s = a\frac{1-e^2}{1+e\cos(\phi)}\sqrt{{\cos}^{2}(\phi){\sin}^{2}(i) + {\sin}^{2}(\phi)}
\label{eq:septrans}
\end{equation}
where $s$ is the projected separation in AU.  When distributed over all values of $i$ ($0^{\circ} \rightarrow 90^{\circ}$), $\phi$ ($0 \rightarrow 2\pi$), and $e$ ($0 \rightarrow 1$) this transforms a single value of $a$ to a range in $s$.  Figure \ref{fig:sepcomp} displays a comparison of the original and the transformed separation distribution models for both the symmetric and asymmetric Gaussian models.

\subsection{Mass Ratio Distribution}

Recall Figure \ref{fig:km2}, which displays the known mass ratio distributions for G, K, early-M stars, and the currently known ultracool binary systems.  Also recall that the mass ratios of G and K binary systems are widely distributed, with a peak near $q = 0.3$.  This begins to change as one looks at lower mass primaries.  Binaries among the early-M dwarfs and the local solar neighborhood, which is dominated by M dwarfs \citep{rg97}, have a mostly-flat mass ratio distribution  Therefore, there seems to be a general evolution toward high mass ratio binary systems with decreasing primary mass.  The ultracool dwarf mass ratio distribution appears to continue this trend.

A power law is used to describe the mass distribution of brown dwarf binary systems.  However, for the binary distribution it is not the mass that is interesting, but the mass ratio, $q = M_2/M_1$.  So, the mass ratio distribution is assumed to follow a power law, $q^{\gamma}$, where $\gamma$ is the power law index.  This power-law is assumed to be constant and does not allow for the roll over or cutoff of the binary mass ratio distribution.  Thus it may overestimate the number of small mass ratio companions..  The index $\gamma$ is not exactly similar to the field or young cluster IMF power-law index $\alpha$, because $\gamma$ is for the combined binary system mass ratio not an individual mass.  Thus, if both the primary and secondary are drawn from a similar IMF then the slope of the mass ratio function ($\gamma$) will be close to 0.  As stated earlier, this is true for intermediate mass G stars.  However, if $\gamma$ is found to be close to zero than it can be said that the low mass binary distribution is drawn from a similar mass function as that of single low-mass stars, (${\sim}-0.3$, A05).  

The overall normalization of the models is factored in at this point.  The normalization, $N$, is set to be the overall binary fraction, the percentage of ultracool dwarfs that have lower mass companions.  When combined with the above power-law in $q$ and a minimum $q$ value of 0.02, the following form is obtained:
\begin{equation}
P(q|N,\gamma){\Delta}q = q^{\gamma}{\Delta}q{\times}N/\sum_{q = 0.02}^{1.0} q^{\gamma}
\label{eq:md}
\end{equation}
where ${\Delta}q$ is the bin size in mass ratio, and $N/\sum_{q = 0.02}^{1.0} q^{\gamma}$ is the normalization.  The left panel of Figure \ref{fig:masscomp} displays model mass ratio distributions for two values of $\gamma$, $-1$ and $+1$.  The $\gamma = -1$ model predicts relatively few equal mass systems and an increasing number of unequal mass systems.  However, for $\gamma = +1$ the distribution is opposite that of $\gamma = -1$, with more equal mass systems than unequal.

These models then need to be transformed to the observable, the magnitude difference of the secondary from the primary at a particular band pass.  The techniques used to transform these models are nearly identical to those used in \citet{a03,a05}.  I use the \citet{bur} ultracool dwarf evolutionary models to provide a link between mass, bolometric luminosity, effective temperature, and age.  However, what is measured is not the bolometric luminosity or effective surface temperature but a magnitude in a particular bandpass.  As in A05, I use the \citet{golim} bolometric corrections to transform the Burrows models into the observable magnitudes from the bolometric.  The bolometric corrections are assigned based on the effective temperatures derived in \citet{golim}.  The main difference between A05 and this work is that I am now concerned with a {\it mass ratio} distribution and not plain mass.  To determine these ratios the transformation requires information on the primary stars.  As stated in A05, each late-M, L, and T spectral type has a unique distribution in mass and age.  This is important because the spectral type of each potential primary is known.  With the range of spectral types sampled in a particular survey, a distribution of mass and age for each primary type and the corresponding range of mass and age for possible secondaries can be constructed.  

Since M6 is the earliest spectral type considered in this analysis, a maximum mass of $0.15~M_{\odot}$ is considered.  The minimum mass is a bit trickier to determine.  Substellar objects have a mass-age degeneracy.  Put simply, as a brown dwarf of a particular mass ages it also cools, thus its spectral type changes.  The latest spectral type in this analysis is a T8.  To take into account this floor to the analysis, I take the absolute magnitude of T8s in the different band passes.  This maximum magnitude translates into a range of masses depending on the age.  As in A05, I assume a flat age distribution.  Thus the mass-age degeneracy of brown dwarfs is accounted for and is built into the model companion distributions.

The right-hand panel of Figure \ref{fig:masscomp} displays the mass ratio distributions transformed to ${\Delta}m_K$ for $\gamma$ equal to $-1$ and $+1$.  For $\gamma = -1$, the structure in the resultant distribution is similar to the luminosity functions generated in A05 Figures 2 \& 3.  This is because, with a power law index of $-1$, the distribution of companions will follow a similar form as single field dwarfs do because that index yields an increasing number of objects with small q.  However, the $\gamma = +1$ distribution is strikingly different.  This model predicts very few unequal mass systems (large ${\Delta}$ magnitudes), and results in a solitary peak at ${\Delta}m_K = 0$, or equal mass systems.  These models are combined with the separation distribution models from the previous subsection to form a complete set of companion distribution models for the Bayesian analysis of the companion distribution.

\section{Bayesian Analysis}
\label{sec:ba}

Before I can proceed with my statistical analysis of the ultracool binary surveys of Section \ref{sec:survey} to determine the underlying companion distribution, we need an understanding of my statistical methodology.  This section provides a brief tutorial on the Bayesian method and how I apply it to constrain the ultracool binary distribution models.  The methods described here are similar to those in A05.

\subsection{Bayesian Core Algorithm}

The core of this method is Bayes' rule:
\begin{equation}
P(\{\theta\}|\{O\}) \propto P(\{O\}|\{\theta\}){\times}P(\{\theta\})
\label{eq:brf}
\end{equation}
where $\theta$ is the model, $\{O\}$ is the set of observed data, $P(\{\theta\}|\{O\})$, the {\it posterior distribution}, is the probability of the model given the data, $P(\{O\}|\{\theta\})$, the {\it likelihood function}, is the likelihood of the data given the model, and $P(\{\theta\})$, the {\it prior distribution}, is the initial probability of the model \citep{siv}.  The output posterior distribution provides a wealth of information on the model parameters, from the best fit parameter set to correlations between parameters.  Using Bayes' rule I calculate the probability that these data would have been measured given a hypothesized model (the likelihood function).  However, I wish to know the probability that a hypothesized model is true given the measured data (the posterior distribution).  The power of Bayes' rule lies in the simple relation of these two quantities.

The specification of a prior distribution remains the most controversial aspect of Bayesian analyzes and must be considered carefully.  The prior folds previous observational and theoretical evidence into the analysis in more than one manner.  Prior information can also be included in the functional forms taken by the models, e.g.\ the Gaussian shape of the separation distribution.  One technique uses the posterior distribution from a previous analysis.  This enables the same analysis to be performed in light of new and improved data.  In this way, one can iterate over multiple data sets, thereby incorporating them into the analysis of a single set of models to provide one unified result.  This is ideal for the study of the ultracool dwarf companion distribution because there is no single data set for all such dwarfs, and new data are continually made available.  Priors can also be constructed if no previous knowledge of the problem exists.  In this case the prior distribution should not impart a bias on any parameter value under consideration.  These types of priors fall under the broad heading of conjugate distributions.  Since there has been relatively little work done on the statistical analysis of the ultracool dwarf companion distribution, the initial prior distributions used here will be constructed using conjugate distributions.  What little prior information there is went into the selection of the forms of the model distributions.

\subsection{Basic Problem and Setup}

The Bayesian analysis of the companion distribution is carried out in the same way as it was for the field mass function in A05.  A Poisson form is chosen as the likelihood function.  Recall the general form of the Poisson likelihood function: 
\begin{equation}
R^{N_{det}}{\times}e^{-N_{obs}{\times}R}
\end{equation}
In this case, the model observations, as derived in Section \ref{sec:cmod}, include $R$, the number of companions per star as a function of ${\Delta}m$ and $\log(s)$.  $R$ is defined as $R(\{O\},\{{\theta}\})$, where $\{O\}$ represents the observable space (${\Delta}m,\log(s))$, and $\{{\theta}\}$ is the vector of model parameters $(N,{\gamma},\log(a_0),{\sigma}_a)$ as defined previously in Section \ref{sec:cmod}.  

$N_{det}$ and $N_{obs}$ must also be defined and constructed.  They are simply two dimensional distributions of the number of detections, $N_{det}(\{O\})$, and the number of times a bin in observable space is sampled, $N_{obs}(\{O\})$.  $N_{obs}(\{O\})$ also describes the window function, in other words, a measure of the areas of observational space covered and the number of observations each area has with respect to the others, see Section 2.2.  With all three components of the Poisson likelihood function defined, it takes on the following form at one point ($\{O\}_{i,j}$) in observable space:
\begin{equation}
P(\{O\}_{i,j}|\{{\theta}\}) \propto R(\{O\}_{i,j}|\{{\theta}\})^{N_{det}(\{O\}_{i,j})}{\times}e^{-N_{obs}(\{O\}_{i,j}){\times}R(\{O\}_{i,j}|\{{\theta}\})}
\label{eq:clike}
\end{equation}
where the subscript $i$ represents the $i$th bin is $\log(s)$, projected separation, and $j$ the $j$th bin in ${\Delta}m$, delta magnitude.

The priors on the model companion distributions must also be determined.  The shape the prior takes depends on the functional form of the models.  What previous knowledge of the ultracool companion distribution there is has gone into the derivation of the functional forms chosen in Section \ref{sec:cmod}.  I wish to fully test the entire range of parameters permitted by these models, thus, I will choose the conjugate priors associated with the model forms I have assumed.  The conjugate distributions for the parameters of a Gaussian distribution are well-established \citep{siv}.  In this case, the prior on the lognormal center separation parameter, $log(a_0)$, is flat or constant.  The width, ${\sigma}_a$, is a scale parameter.  A maximum entropy argument can be used to determine the prior distribution on ${\sigma}_a$.  To do so, the constraints to which the distribution on ${\sigma}_a$ is subject must be defined.  The only constraint on the probability distribution of ${\sigma}_a$ is an invariance to changes in scale; i.e., the units of distance can be changed from AU to centimeters with no effect on the outcome.  This means that ${\sigma}_a$ is a scale parameter, and the most `ignorant' prior distribution is given by $P({\sigma}_a) \propto 1/{\sigma}_a$ \citep{siv}.  The normalization, $N$, is also a scale parameter; it is required to be independent of the units of the calculation.  The power-law index, $\gamma$, of the mass ratio distribution is unconstrained and does not depend on the scale, so it is given a flat prior.  The complete prior distribution is given as follows:
\begin{equation}
P(\{{\theta}\}) \propto \frac{1}{N}{\times}\frac{1}{\sigma_a}
\label{eq:cprior}
\end{equation}

Equations \ref{eq:clike} and \ref{eq:cprior} together form the RHS of the unnormalized Bayes' rule (Equation \ref{eq:brf}).  This form is only true for one point in observable space, $\{O\}_{i,j}$.  However, it can easily be manipulated to be both easier to calculate and cover the entire observable space.  Assuming that the observation of the companion rate in an observed bin, $\{O\}_{i,j}$, is independent of the measurement in any other bin, the likelihoods and priors at each observed point can be multiplied to obtain the following form:
\begin{equation}
P(\{{\theta}\}|\{O\}) \propto P(\{{\theta}\}){\times}\prod_{i,j}P(\{O\}|\{{\theta}\})_{i,j}
\label{eq:rhs}
\end{equation}
where the subscripts are the same as in Equation \ref{eq:clike}.  The prior distribution is outside the product because it is a function of the model parameters and not the observable dimensions.  The calculation becomes easier, in the same way as the field mass function analysis in A05, if the natural logarithm is taken of both sides of Equation \ref{eq:rhs}.  It takes on the following form:
\begin{eqnarray}
\nonumber ln(P(\{{\theta}\}|\{O\})) = ln(P(\{{\theta}\})) + \sum_{i,j} N_{det_{i,j}}(\{O\}){\times}ln(R(\{O\}|\{{\theta}\})_{i,j}) - \\
N_{{obs}_{i,j}}(\{O\}){\times}R(\{O\}|\{{\theta}\})_{i,j} + constant
\label{eq:cfinal}
\end{eqnarray}
where the sum is over the two observable dimensions, ${\Delta}m_i$ and $\log(s)_j$.  This is the final form of the likelihood function, and it will be used in Section \ref{sec:res} to constrain the binary distributions models.

\section{Results}
\label{sec:res}

To perform the Bayesian calculations, the models, described in Section \ref{sec:cmod}, and the data, from Section \ref{sec:survey}, are fed into Equation \ref{eq:cfinal}.  The output posterior distribution is four dimensional (${\log}(a_o),\sigma_a,N,\gamma$).  As a result, it is impossible to display the entire distribution, so it is marginalized over different sets of parameters.  This process simply means the collapse of the posterior distribution along different parameter axes.  The marginalized distributions of each individual parameter are displayed in Figures \ref{fig:jo} and \ref{fig:jo2}, for the symmetric and asymmetric Gaussian models respectively.  The best fit parameter values, with $1\sigma$ uncertainties, are shown in Table 1.  Each individual parameter distribution is well-behaved and exhibits a clear peak.  

\subsection{Model Parameter Results}

The parameter values returned by the two different models are very similar (Table 1).  This is not surprising because the region where they differ (separations $>$10~AU) has very few detections.  However, we should recall that many of the detections that have been made at these wide separations have been excluded from this analysis.  Both parameter sets are consistent in finding a peak to the separation distribution near 7~AU, an overall binary frequency in the low 20\%'s, and a steep mass ratio spectrum (${\gamma}\sim1.8$).  The main difference lies in the Gaussian width parameter, $\sigma_a$.  The best fit values are quite similar but the $1\sigma$ uncertainties are different by a factor of two, with the asymmetric model having the larger uncertainty.  The wider variation in $\sigma_a$ in the asymmetric Gaussian model allows the possibility of a significantly broader distribution to smaller separations.  This would result in a large population of, currently, undetected spectroscopic binaries, as predicted in \citet{jm05}.

The tails of the $N$ and $\log(a_0)$ distributions are correlated.  Figure \ref{fig:sono} displays the two-dimensional, marginalized posterior distribution of those two parameters for the symmetric Gaussian model.  The tail to small values of $\log(a_0)$ corresponds to the tail of the companion rate to larger values.  This is because no current imaging survey is sensitive to companions at separations less than 1-2~AU (see Figure \ref{fig:winfunc}).  The Bayesian algorithm allowed parameter values such that there are a large number of companions at very small separations, where no observations exist to constrain the overall binary fraction.  This reconstructs the observed data, but with lower likelihood (you must move 2-3$\sigma$ from the best fit values in Figure \ref{fig:sono}) than the best fit model parameter set.  This tail does permit the existence of significant population of very closely separated companions, as predicted by \citet{jm05}.  Further studies must be carried out to determine the frequency at separations less than a few tenths of an AU.  While the best method for this is a spectroscopic survey for either radial velocity variations or double-lined systems, there are other techniques that may be used.  These methods are widely varying and include sources with known parallaxes which are over luminous, astrometric wobbles, spectral inconsistencies \citep{cruz04}, and over luminous young cluster members.

A further intriguing difference between the results of the two different models can be seen in panel c of Figure \ref{fig:jo2}.  There is an auxiliary peak at $\sigma_a = 0.2$, which suggests there is a secondary solution.  This solution is quite different than the two main solutions, see best fit parameters values in Table 1.  The uncertainties, while lower for the separation parameters, are considerably higher for the mass ratio distribution parameters than the two main solutions.  This solution is for a very narrow separation distribution, but a considerably higher binary frequency (28\%).  The higher binary rate is necessary to produce the correct number of companions in the wings of the separation distribution to compensate for the narrower value of $\sigma_a$.  Finally, the relative likelihood of the secondary solution compared to the primary solution is significantly different.  These likelihoods are the result of the calculation of Equation \ref{eq:cfinal}, and the difference is $-100$.  Recall, that the likelihoods are given by the natural logarithm of the probability.  The actual difference in probability between the primary and secondary solutions is $e^{-100}$, so the secondary is considerably less likely.

\subsection{Mass Ratio Distribution}

One of the most significant results is seen in panel d of Figures \ref{fig:jo} and \ref{fig:jo2}, on $\gamma$, and is reproduced in Figure \ref{fig:gam}.  As noted earlier, there was a selection bias towards ultracool dwarf companions to near-equal mass, tightly bound systems.  Using the correction factor from Equation \ref{eq:selbias} and the Bayesian analysis, I can determine if this is just the result of the bias, or if it is a real feature of the ultracool dwarf companion distribution.

The most likely value of $\gamma$ for the companion mass function is ${\sim}1.8\pm0.6$, to $68\%$ confidence.  While this parameter is not well constrained it is different from the Salpeter slope ($-2.35$) for very massive stars to $7\sigma$, the index for G, K, and early M stars ($-1.05 \pm 0.15$) \citep{rg97} to $5\sigma$, and the best fit model for field ultracool dwarfs from A05 ($-0.3{\pm}1.0$) to $3.5\sigma$.  Values of $\gamma$ below 0.0 are ruled out to the 99.95\% confidence level.  \citet{burg06m} finds a much larger value of $\gamma$, $4.2\pm1.0$.  This steeper value may be due to the lack of M dwarf primaries in their analysis, which concentrated on L and T dwarfs.  The large number of late-M dwarfs in this work may have softened the slope of the mass ratio distribution.  Thus, it may be that the mass ratio distribution continues to steepen even further with decreasing primary mass.  Overall, the values of $\gamma$ calculated here and in \citet{burg06m} imply that the mass distribution of brown dwarf companions is fundamentally different from that of isolated ultracool dwarfs, marked by the slashed rectangle in Figure \ref{fig:gam}.  This is also a different companion distribution from higher mass G and K stars, but continues the trend of lower-mass primaries (M and later) having larger fractions of $q = 1$ companions.  For comparison, Figure \ref{fig:km} displays the mass ratio distributions plotted in Figure \ref{fig:km2} with the best fit companion mass ratio model.

\subsection{Separation Distribution}

The most significant separation distribution result of this analysis is that the dearth of wide companions is a real feature of the ultracool binary distribution.  As seen in panels b and c of Figures \ref{fig:jo} and \ref{fig:jo2}, the posterior distributions for the separation distribution is well-constrained with a Gaussian center of ${\sim}7.2~AU^{+1.1~AU}_{-1.7~AU}$.  Figure \ref{fig:kc2} is the same as Figure \ref{fig:kc} but with the best fit separation distribution displayed as well.  The binaries that appear at separations $>{\sim}30$~AU in the \citet{burg07} data are all companions associated with primaries not included here and thus were excluded from this analysis.   However, the overall frequency of wide companions ($>$15~AU) predicted by the best fit models is ${\sim}1-2\%$ (Table 1).  This is still consistent with the relative fraction of wide companions if the recently detected wide, young systems are included.  It should be noted that wide ultracool companions to G, K, and early-M stars is a different matter.

\subsection{Missing Companions?}

For the two primary best fit solutions, I determined the binary fraction missed by the surveys used in this work as a function of separation.  Figure \ref{fig:miss} displays the results.  Note that the fit for the asymmetric Gaussian produces a similar number of missing close companions as the symmetric Gaussian.  This is because there is sufficient sensitivity in the current surveys to see a downturn in the binary frequency to smaller separations.  However, this is a lower limit to the missing fraction.  The surveys considered here do not have any sensitivity to very close companions and these results may simply be caused by the lack observations.

The vast majority of the missed binary fraction is for separations less than 10~AU.  These missing companions are low-q systems that the surveys were not sensitive to and tight spectroscopic systems with a range of q's.  The overall fraction missing is $6\%-7\%$.  This means that an additional $\sim$6\% of ultracool primaries have undetected companions.  This is perfectly consistent with observational results.  The observed binary fraction is $\sim$15\%, so when the missing fraction is added to the observed we recovered the best fit values for the ultracool dwarf companion frequency.  Thus, according to my predictions, current observational efforts have recovered 60\% to 75\% of ultracool dwarf binaries.

I have also tested the possibility that there is a large reservoir of low q binary systems to which the surveys used in this work were not sensitive.  In the analysis of main sequence spectroscopic binaries, \citet{halb} and \citet{gold} show that the mass ratio distribution is bi-modal.  There is a peak in the distribution near q of 0.8 to 1.0, similar to that found here, as well as a broad peak around a q of 0.2.  

To simulate this sort of distribution, I determined the distribution of delta magnitudes that corresponds to a broad peak around q of 0.2.  This is generally reconstructed, for the $K$-band, as a Gaussian distribution centered at ${\Delta}M_K$ of 6 with a width of 4 magnitudes.  To test the detectability of this low q distribution, I created a series of ${\Delta}M_K$ distributions.  I started with the best fit symmetric Gaussian model and simply added the new low q extension.  I normalized this extension to different additional binary frequencies, $N_a$, in the range of 5\% - 45\%.  I chose this range to correspond to, at a minimum, the 1~$\sigma$ upper limit on the value of $N$ and, at the maximum, to give near parity between the high q and low q peaks, as is seen in \citet{halb} and \citet{gold}.  The new parameter $N_a$ is just an additional binary frequency, thus a model with $N_a = 10\%$ has a total binary frequency of 30\% (recall the baseline rate of the best fit symmetric Gaussian model of 20\%).  Figure \ref{fig:lqt} displays an example of this modified delta K magnitude distribution.  Here we see the sharp peak at small delta magnitudes from the best fit symmetric Gaussian and the new broad distribution of low q companions.

I determined the number of companions that would have been detected at ${\Delta}M_K > 2$ in the surveys conducted in the K-band \citep{k99,close03,ns03,ns05,a07} by simply multiplying the window function by the revised model.  These numbers were then compared both to the number actually detected (2) and the number predicted by the best fit model (1.2).  The number derived for each new model delta magnitude model is displayed in Figure \ref{fig:lqt2}.  At the 1~$\sigma$ level only the lowest value of $N_a$ is barely consistent with the observations, which is also consistent with the uncertainties on the fit of $N$.  Thus, it is not likely that there is a significant population of low-q companions that have not been detected.

\subsection{Reliability of the Results}

A goodness of fit test is used on the best fit models to see if the companion model form, Equations \ref{eq:sepinit}, \ref{eq:sepinit2}, \& \ref{eq:md}, accurately represents the data.  A standard Monte Carlo technique is used.  Five hundred realizations of each best fit companion distribution model are randomly drawn.  These realizations are constrained to have the same number of detected companions as the real data sets.  The likelihood of the best fit to each fake data set is calculated by running them through the Bayesian algorithm (Equation \ref{eq:cfinal}).  If the best fit model truly reflects the real data then the likelihood of the best fit of the observed data should fall near the peak of the fake maximum likelihoods.  Figure \ref{fig:monte} displays the results, and the best fit model does lie well within the likelihoods calculated from the fake data sets.  However, both the symmetric and asymmetric Gaussian models produces similarly good fits, so we cannot, at this time, differentiate between the two.  Thus, both models can reconstruct the observed data, the main difference being that the asymmetric Gaussian model may provide for a substantial population of undiscovered close binary systems.

I also conducted a similar exercise for the secondary asymmetric Gaussian solution.  As can be seen in Figure \ref{fig:monte2}, this fit is not good.  The peak of the fake likelihoods lies to the right of the best fit values, so that model does not provide a good representation of the data.  Overall, the ultracool dwarf companion models do a reasonable job of reconstructing the data and both must be considered valid at this time.

\section{Discussion and Conclusions}
\label{sec:conc}

While the constraints placed by my Bayesian analysis are not tight they do reinforce the significance of three trends already noted in the ultracool dwarf binary distribution.  First, {\it the peak at $q \sim 1$ is a real feature}.  The sensitivity of the high-resolution and wide-field imaging surveys used in this analysis, while falling to smaller mass ratios, does not fall off significantly until q's of about 0.7-0.5.  Therefore the statistical significance of the peak towards high mass ratio systems is genuine as seen in the posterior distribution for $\gamma$ (Figure \ref{fig:gam}).  

Second, {\it the severe lack of companions at separations greater than ${\sim}15-20$~AU}.  At this time, there are only a handful of confirmed field ultracool binaries with separation greater than 20~AU \citep{lu04,close06,chau04,luh06,bill05,art07}.  The best fit models generated here predict wide binary fractions of ${\sim}1\%$.  This is a very robust result.  Additionally, I tested the idea of an undetected spike in the mass ratio distribution at low ($q{\sim}0.2$), and it was shown that a significant population was not consistent with the observations (Figures \ref{fig:lqt} and \ref{fig:lqt2}).  Between all of the imaging surveys considered in this paper there were 361 observations for companions with separations between 20~AU and $\sim$1000~AU and {\it zero} detections.  However, recent work by \citet{bill05} using DENIS objects has uncovered one very wide candidate companions in a survey of 250 potential primaries.  Therefore, the rough binary fraction of wide companions is $\sim$1\%, which is in good agreement with my predicted results from the symmetric Gaussian.  

There have been four detections of wide companions to young ultracool dwarfs \citep{luh04,luh05,luh06,chau04,chau05,close06}.  The appropriate selection effects for these surveys are not well quantified making it difficult if not impossible to determine a binary frequency, thus precluding the addition of these results to my Bayesian analysis.  Nevertheless, the greater prevalence of wide and low mass ratio systems among young objects is intriguing.  As has been discussed in \citet{burg07}, this is based on very small number statistics (8 binaries) and, therefore, requires a significant amount of further study to verify.  However, the overall low rate from \citet{bill05} of $\sim$1\% is perfectly consistent with the outcome of my Bayesian analysis and the Gaussian separation distribution models which predict a similar frequency of wide companions.

Third, {\it the decreasing of the binary fraction with decreasing spectral type primaries.}  I predict an overall ultracool dwarf multiplicity of ${\sim}20$\%.  This continues the decreasing trend from G stars, ${\sim}60\%$, to M stars, ${\sim}30\%-40\%$, to my result.  My results are also consistent with the most recent empirical estimates in the literature, including the $\sim$24\% overall binary frequency of \citet{inr06} and the $20\%-25\%$ of \citet{br06}.  Clearly, on overall observational picture is taking shape, which, as my Bayesian analysis demonstrated, has great statistical significance.

There are implications for the formation mechanism models based on the conclusions of this work.  The two empirical facts that must be accounted for are the steep mass ratio distribution that peaks at $q=1$ and the $\sim$1\% frequency of wide systems.  Unfortunately, few of the numerical simulations produce concrete predictions for the distribution of binary systems.  Some notable exceptions are the hierarchical triple system decay models by \citet{um}, which produces a separation distribution with a peak at a projected separation of 3~AU, as well as sharp declines to both tighter and wider systems.  This qualitatively agrees with the best fit solutions from my analysis.  The peak of the distribution is roughly half of what I find (3~AU instead of 7~AU).  However, their distribution reproduces the few tighter and wider binary systems of my best fit models.  Thus, the Umbriet et~al.\ models are in agreement with my analysis of the data.  The work of \citet{ster} on dynamical interactions of small-N clusters was able to qualitatively reproduce the decreasing binary fraction with decreasing primary mass.  This is another result that I confirm in this work.  

One model that has received a great deal of attention in the last several years is that of proto-stellar ejection, as pioneered in \citet{rc01}.  The Bate et~al.\ group \citep{bbb02,bbb03,bb05} has produced some of the most complete simulations of turbulent fragmentation and ejection of brown dwarf embryos.  The binary fractions that they produce are 5\%-8\% for the separation ranges that they probe ($>2.5~AU$).  Their combined work generated $\sim$20 binaries, only one of which was a wide system (20~AU).  To compare these predictions to the results of this work I determined the binary fraction of the best fit models at separations greater than 2.5~AU.  This matches the Bate et~al.\ separation range, and both the symmetric and asymmetric Gaussian models yield binary fractions of $\sim$14\%.  This is a factor of 2-3 higher than predicted by the ejection model simulations.  Thus, the ejection hypothesis of ultracool dwarf formation must, as illustrated by the simulation results compared to those here, be held in doubt as it is unable to reproduce the ultracool binary distribution.  

There are three major observational efforts that need to be undertaken before the observations can be considered complete.  First and foremost is the current lack of large spectroscopic surveys for companions at less than $\sim$1~AU.  This effort is just beginning through the work of \citet{jm05,vj06,br06}.  The results from these preliminary surveys vary significantly, from $32-45\%$ in \citet{jm05} to $26\%\pm10\%$ in \citet{br06}.  \citet{jm05} found their binary frequency through statistical inference of 2-3 radial velocity measurements of many objects, whereas the \citet{br06} found theirs through consistent monitoring of a small nearby sample.  \citet{vj06} examined several young objects in the Chameleon 1 star-forming cloud and did not find a single confirmed companion.  The results of this work are in agreement with the \citet{br06} findings, and demonstrate that the inferred \citet{jm05} binary fractions are likely too high by a factor of 2.  Table 1 shows the predicted binary fractions for tight ($<$1~AU) systems for both the symmetric and asymmetric Gaussian models.  I find that only an additional 3\%-4\% of ultracool dwarfs should have such tight binary systems.  The discrepancy between the \citet{jm05} and the concurring results of this work and \citet{br06} highlights the need for a well-defined volume-complete sample of spectroscopic binary observations.  Such a sample is crucial to constraining this sparsely sampled region of parameter space.  

Second, is a region of parameter space that is nearly unconstrained, wide very low-mass ratio companions.  It is unlikely that there is a significant reservoir of unseen extremely low-mass companions to ultracool dwarfs (Figure \ref{fig:lqt2}), but it must be explored for completeness.  Also note that the majority of the wide companions found thus far have low mass ratios.

Finally, high-resolution imaging surveys need to be improved to reliably reach into separations on the order of a few tenths of an AU.  This is because the sensitivity of the radial velocity searches falls off at this point and the sensitivity of the current imaging surveys begins to fall off at around 3-4~AU.  This leaves an insensitive hole in efforts to observe binary companions to low mass stars and brown dwarfs near where we are currently seeing a peak in the distribution!  This must be resolved to accurately measure the distribution of companions.

I would like to thank my referee, Adam Burgasser, for his varied and insightful comments.  I also thank David Koerner for the original idea that inspired this work and for the many fruitful discussions we have had in the interim, and Neill Reid for his financial and scientific support.  I would also like to thank Adam Burrows for the use of his models and answering questions about them, and David Trilling and Kelle Cruz for many hours of conversation and their support.  Finally, I gratefully thank Gary Bernstein for many helpful discussions on statistics.  I acknowledge support by a grant made under the auspices of  the NASA/NSF NStars initiative, administered by JPL, Pasadena, CA, and support from grant NAG5-11627 from the NASA Long-Term Space Astrophysics program made to Kevin Luhman.

\clearpage

\clearpage

\clearpage

\begin{figure}
\centering
\includegraphics[scale=0.4,angle=-90]{f3a.ps}
\includegraphics[scale=0.4,angle=-90]{f3b.ps}
\caption{Left: Histogram of the distribution in spectral type of the individual target primaries of the surveys described in Section \ref{sec:survey}.  Right: Distance distribution of the surveys: Ground-Based AO \citep{close03,ns03,ns05} (red asterisks), WFPC2 HST \citep{inr01,gizis03,bouy03,burg03} (green circles), Keck (Koerner et~al, in prep) (blue x's), IRTF \citep{a07} (cyan squares), and HST NICMOS \citep{inr06,burg06m} (purple triangles).}
\label{fig:spt}
\end{figure}

\begin{figure}
\centering
\includegraphics[scale=0.3,angle=90]{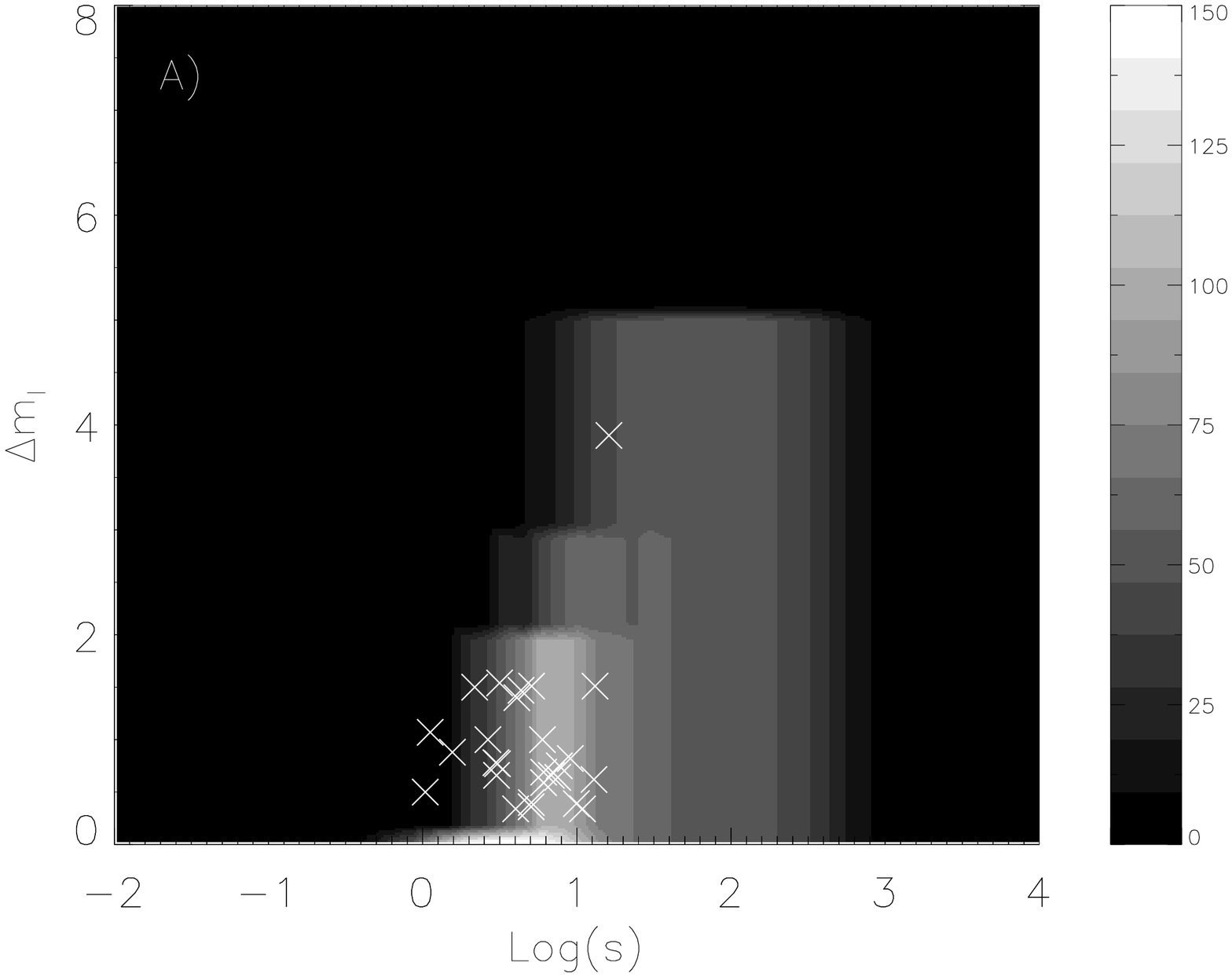}
\includegraphics[scale=0.3,angle=90]{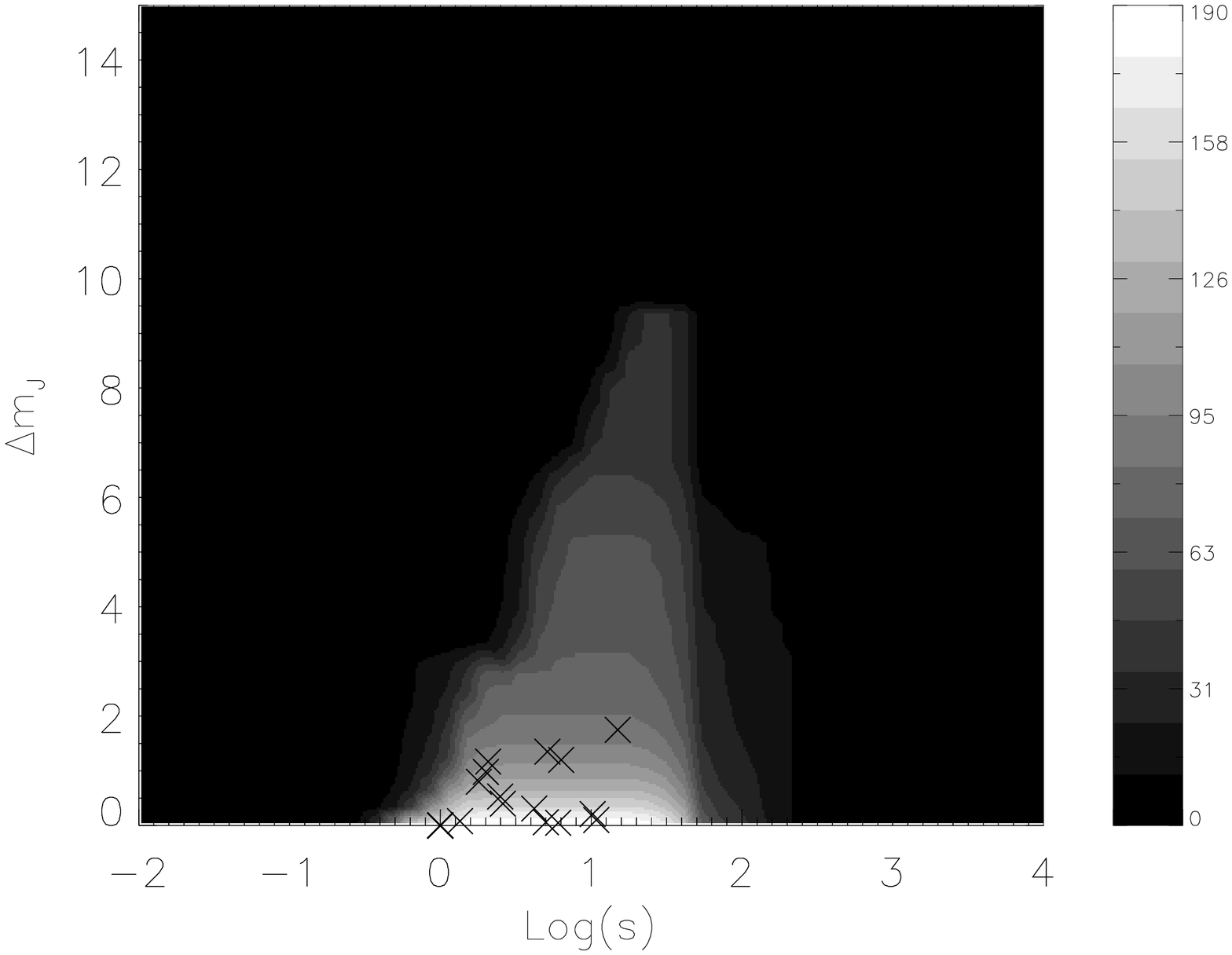}
\includegraphics[scale=0.3,angle=90]{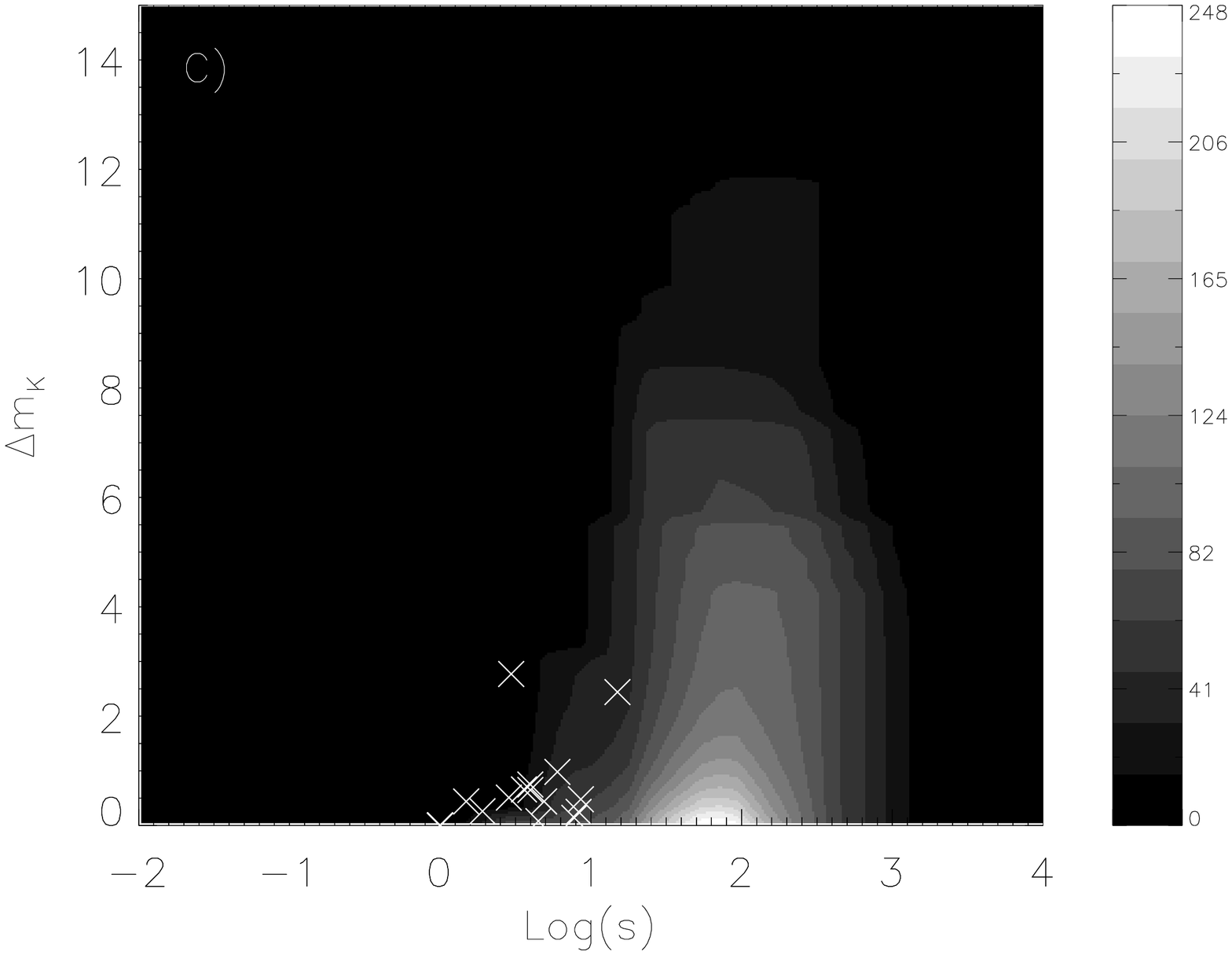}
\caption{The observed window functions for the combined ultracool dwarf companion surveys in delta magnitude and projected separation in AUs.  The top panel represents the observations of the optical HST surveys \citep{bouy03,gizis03,inr01,burg03}, the middle panel the near infrared NICMOS observations \citep{inr06,burg06m}, and the bottom panel all the near-infrared observations in the K-band \citep{close03,ns03,ns05,a07,k99}.  The white (black in the middle panel) crosses in both panels represent the known binaries from the respective surveys.  The scale bars give the number of observations that corresponds to a particular color.}
\label{fig:winfunc}
\end{figure}

\begin{figure}
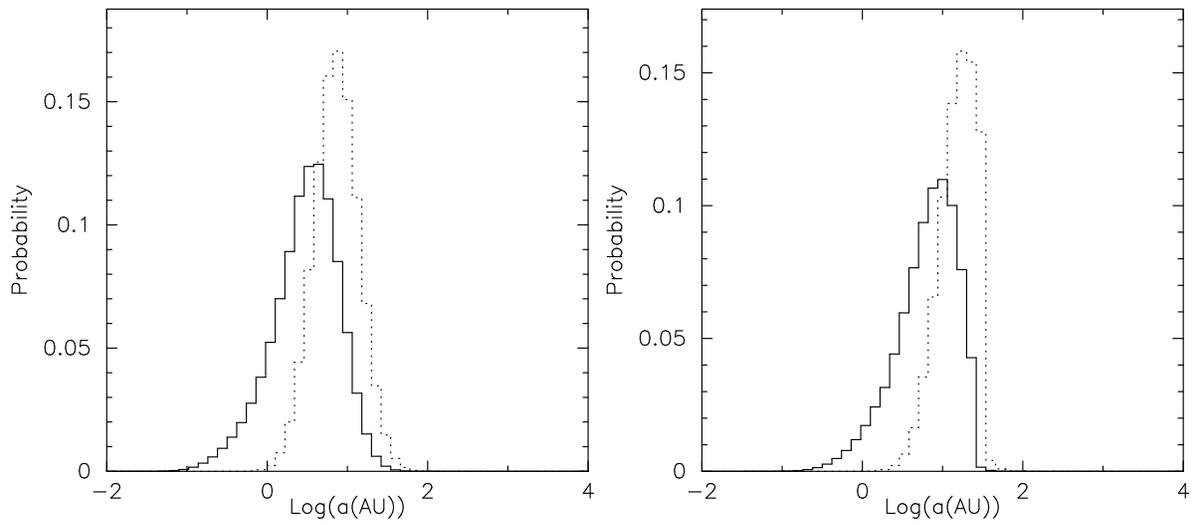

\centering
\includegraphics[scale=0.4,angle=-90]{f5a.ps}
\includegraphics[scale=0.4,angle=-90]{f5b.ps}
\caption{Comparison of the raw separation distribution model in $\log(a)$ (solid histogram) to the transformed projected separation ($\log(s)$) (dotted histogram) for the symmetric Gaussian model (left) and the asymmetric Gaussian model (right).}
\label{fig:sepcomp}
\end{figure}

\clearpage

\begin{figure}
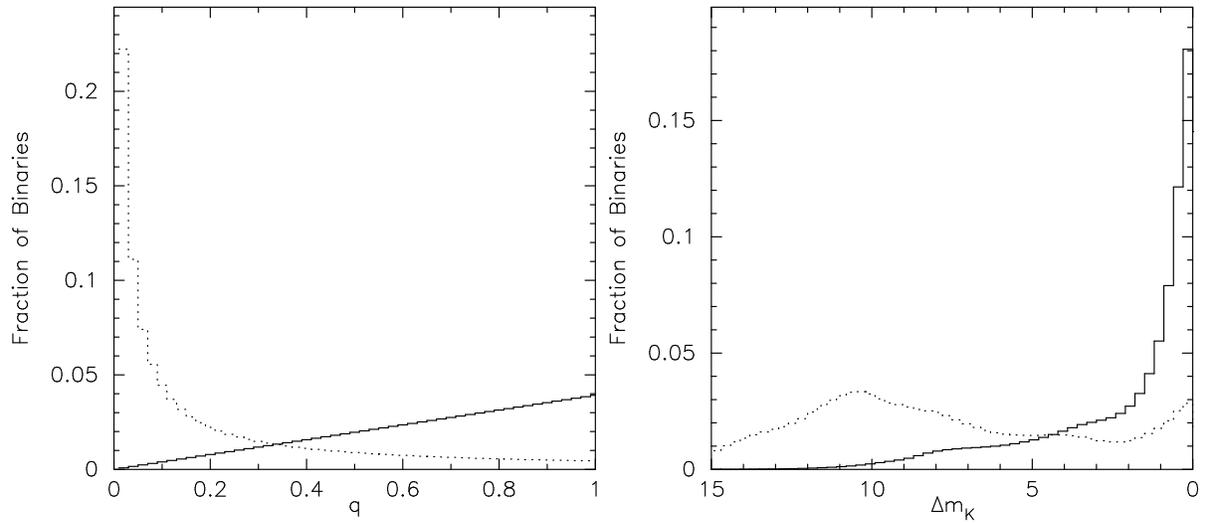

\centering
\includegraphics[scale=0.4,angle=-90]{f6a.ps}
\includegraphics[scale=0.4,angle=-90]{f6b.ps}
\caption{Comparison of the raw mass ratio distribution model (left) and the transformed model in delta magnitude (right) in this case ${\Delta}m_{K}$.  Also compared are two different values of the mass ratio distribution parameter $\gamma$, the power-law index, $\gamma = -1$ (solid lines) and $\gamma = +1$ (dashed lines).}
\label{fig:masscomp}
\end{figure}

\clearpage

\begin{figure}
\centering
\includegraphics[scale=1.,angle=-90]{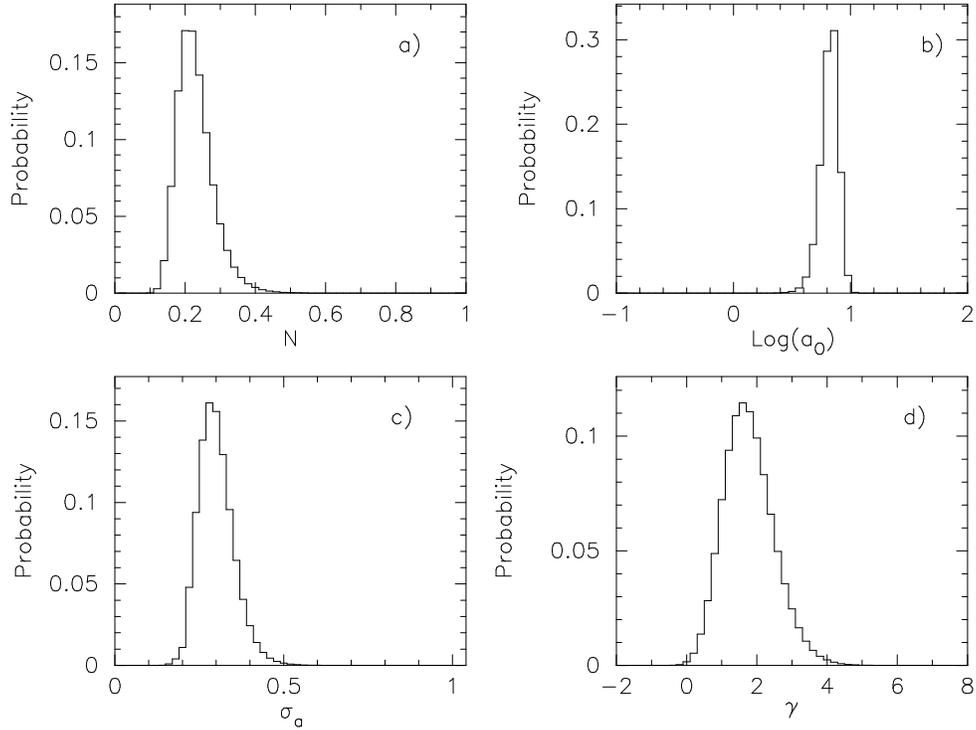}
\caption{Marginalized posterior distributions on the four parameters of the symmetric Gaussian binary companion distribution model: a) The companion rate per star at near equal masses, $N$; b) The center of the Gaussian separation distribution, $\log(a_0)$; c) The width of the Gaussian separation distribution, $\sigma_a$; d) The power-law index of the mass ratio distribution of companions, $\gamma$.  Note the long tail on $N$ to high values and the tail on $\log(a_0)$ to smaller values.  These tails are correlated, as seen in Figure \ref{fig:sono}.}
\label{fig:jo}
\end{figure}

\begin{figure}
\centering
\includegraphics[scale=1.,angle=-90]{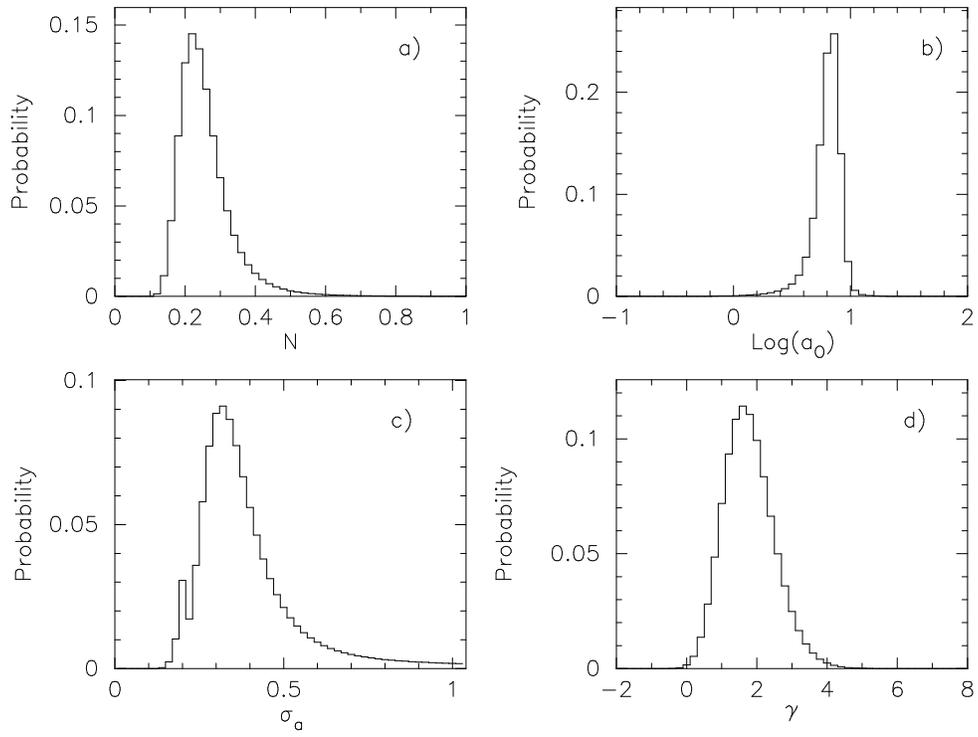}
\caption{Same as Figure \ref{fig:jo} but for the asymmetric Gaussian binary companion distribution model.}
\label{fig:jo2}
\end{figure}

\begin{figure}
\centering
\includegraphics[scale=0.55,angle=90]{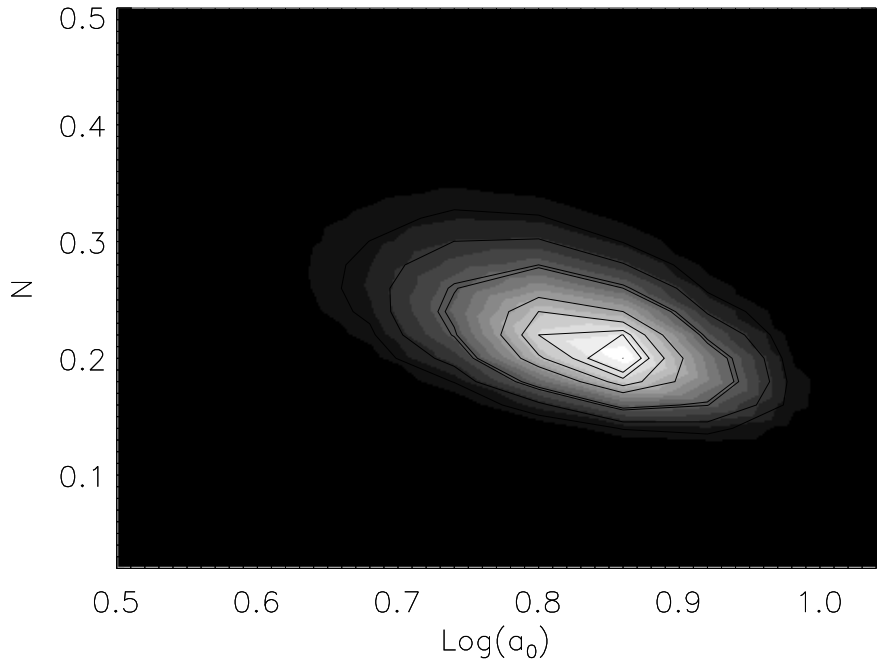}
\caption{Marginalized 2D posterior distribution on the companion fraction, $N$, and the Gaussian center parameter, $\log(a_0)$ for the symmetric Gaussian model.  The contours denote 10\% confidence intervals.  Note how these two parameters are correlated.  A smaller value of $\log(a_0)$ requires a larger value of $N$ in order to  recreate the correct number of detections at larger separations.}
\label{fig:sono}
\end{figure}

\begin{figure}
\centering
\includegraphics[scale=0.55,angle=-90]{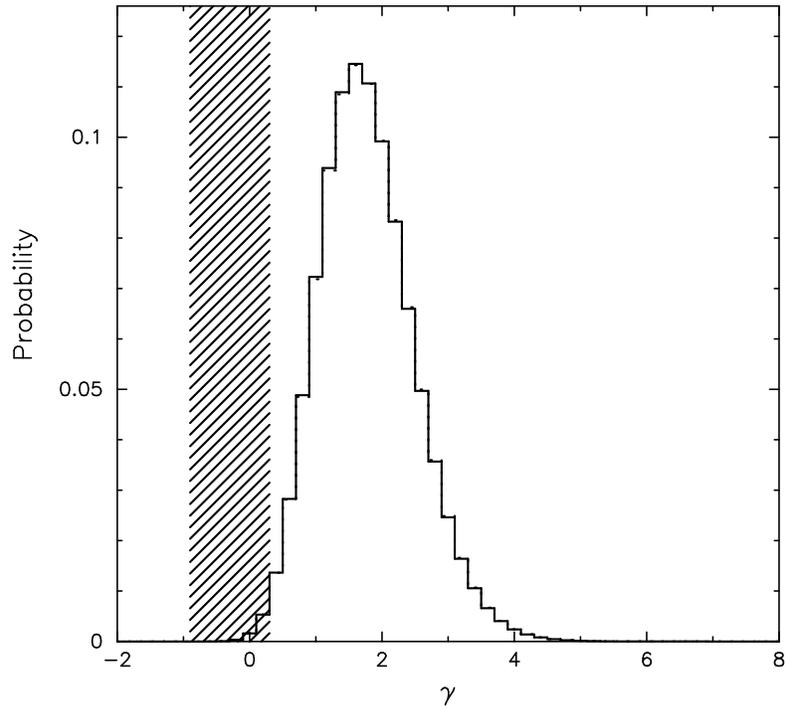}
\caption{Marginalized posterior probability distribution for the power-law index $\gamma$, for the companion mass ratio.  This is reproduced from Figure \ref{fig:jo}d.  The corresponding range of mass function power-law indices derived for field objects in \citet{a05} is displayed as a vertical slashed rectangle.}
\label{fig:gam}
\end{figure}

\begin{figure}
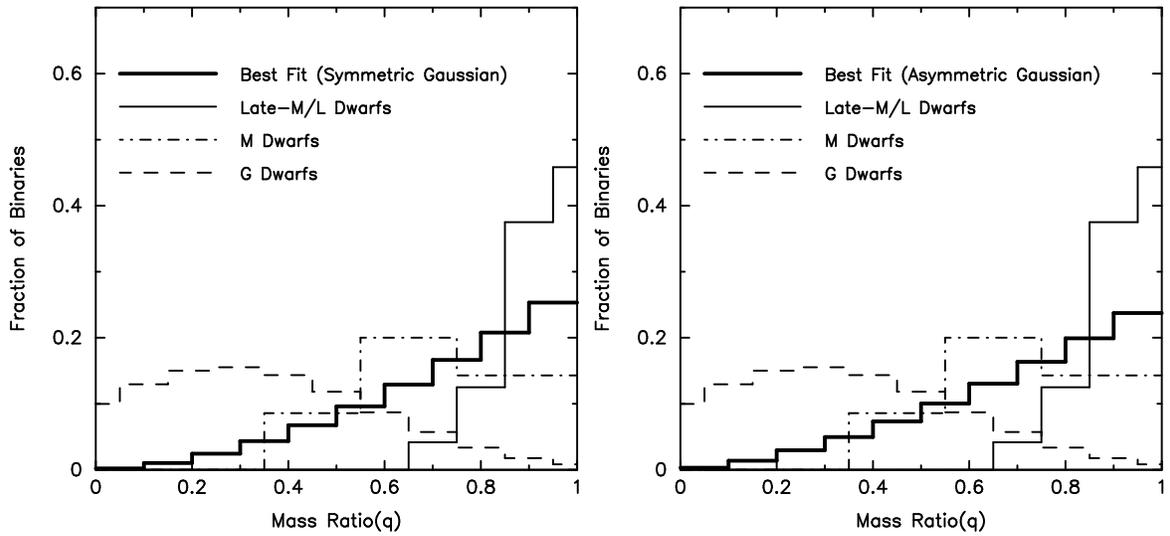

\centering
\includegraphics[scale=0.4,angle=-90]{f11a.ps}
\includegraphics[scale=0.4,angle=-90]{f11b.ps}
\caption{Similar to Figure \ref{fig:km2}, but also displays the Bayesian best fit mass ratio model (thick solid lines) for the symmetric Gaussian (left) and the asymmetric Gaussian (right).  Note that the best fit is generally consistent with the prevalence of near-equal mass companions to ultracool dwarfs.}
\label{fig:km}
\end{figure}

\begin{figure}
\centering
\includegraphics[scale=0.7,angle=-90]{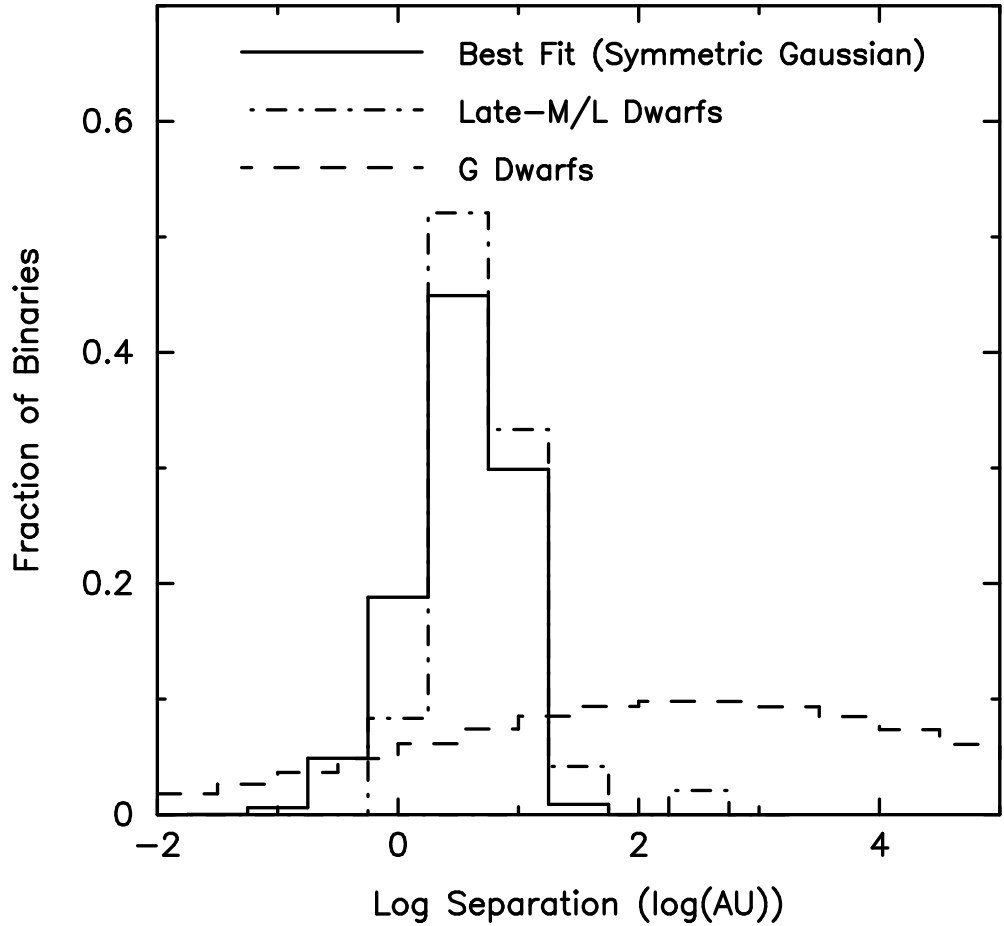}
\caption{Similar to Figure \ref{fig:kc}, but also displays the Bayesian best fit symmetric Gaussian separation distribution model (solid histogram).  Here we see that there is likely a modest population of smaller separation binaries that will be discovered with current and future spectroscopic surveys.}
\label{fig:kc2}
\end{figure}

\begin{figure}
\centering
\includegraphics[scale=0.55,angle=-90]{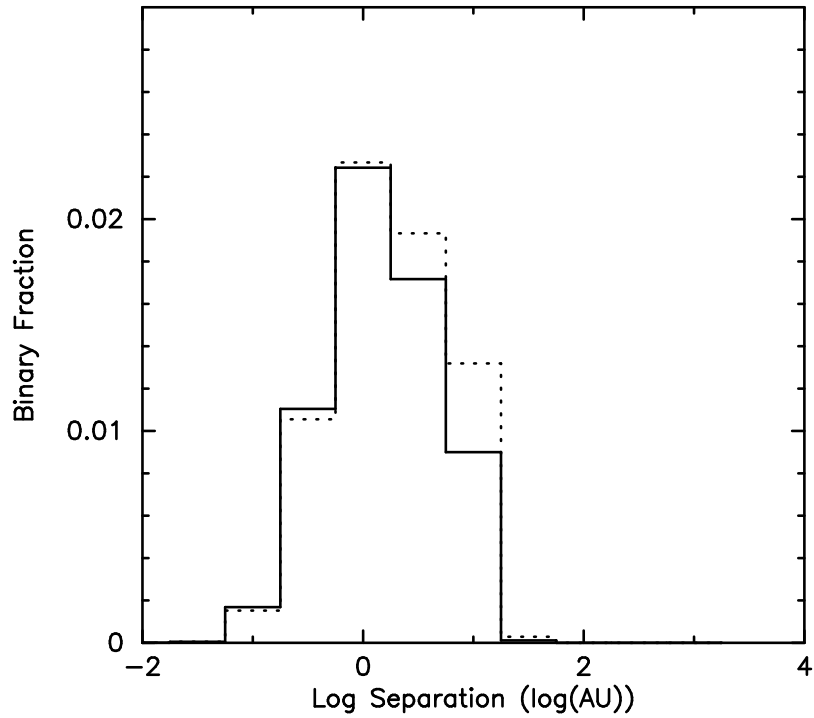}
\caption{Predicted missing binary fraction for the symmetric Gaussian (solid histogram) and the asymmetric Gaussian (dotted histogram) models.  The binary fraction displayed here is the overall, not the relative.  For example, these models predict that an additional 2.2\% of ultracool primaries have undetected companions at separations of 1~AU.}
\label{fig:miss}
\end{figure}

\begin{figure}
\centering
\includegraphics[scale=0.7,angle=-90]{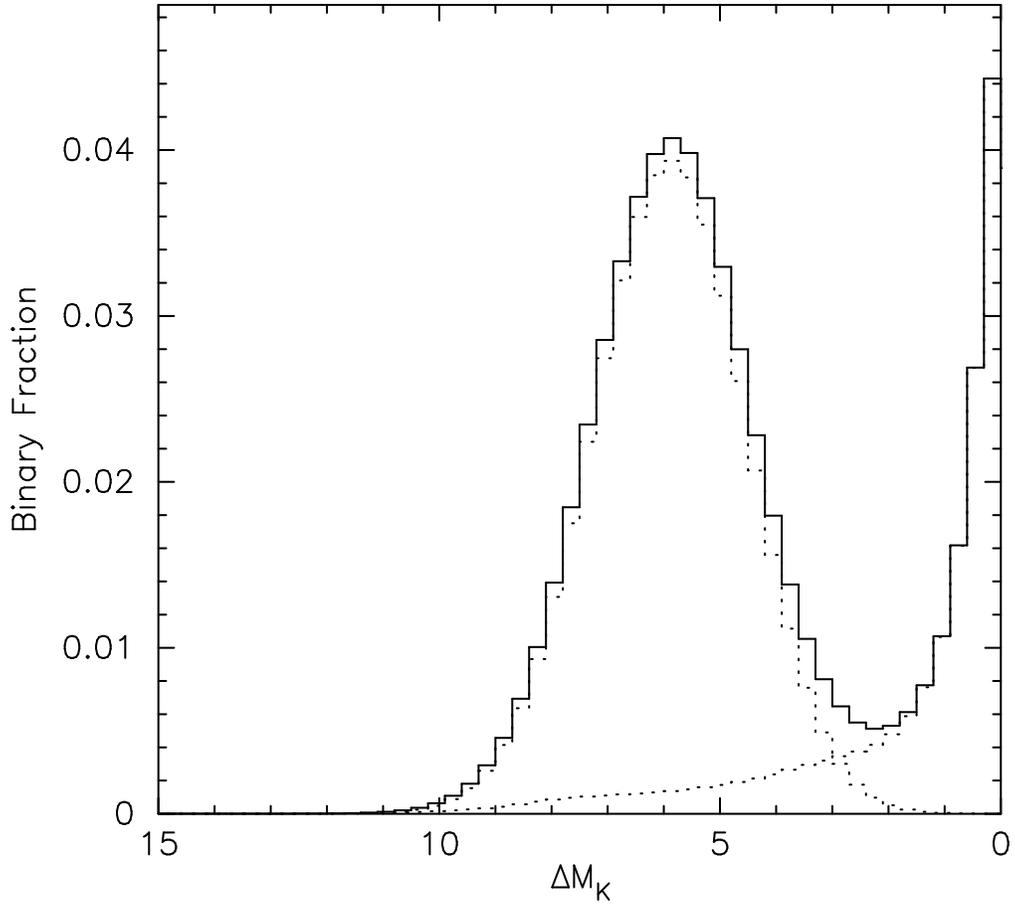}
\caption{Displays the sum of the symmetric Gaussian best fit ${\Delta}M_K$ distribution plus the added low q test distribution.  The test distribution used here is for an added companion fraction of 45\%.  The solid histogram is the sum of the two distributions while the dotted histograms are for the two separately.}
\label{fig:lqt}
\end{figure}

\begin{figure}
\centering
\includegraphics[scale=0.7,angle=-90]{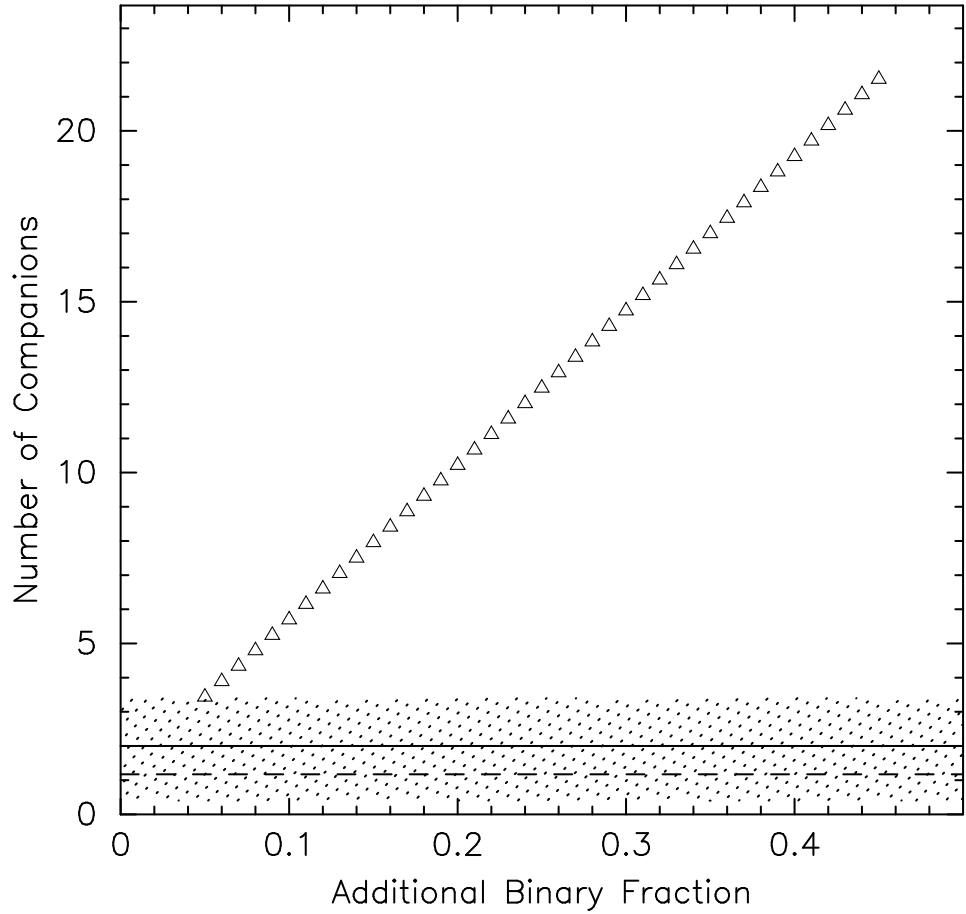}
\caption{Displays the number of companions that would have been detected at ${\Delta}M_K$'s greater than 2 (open triangles) as a function of the added binary frequency of the low q test distributions.  Also shown are the number of companions predicted by the best fit symmetric Gaussian model (dashed line), the number of companions actually detected (solid line), the $1\sigma$ uncertainties of that measurement (dotted rectangle).}
\label{fig:lqt2}
\end{figure}

\clearpage

\begin{figure}
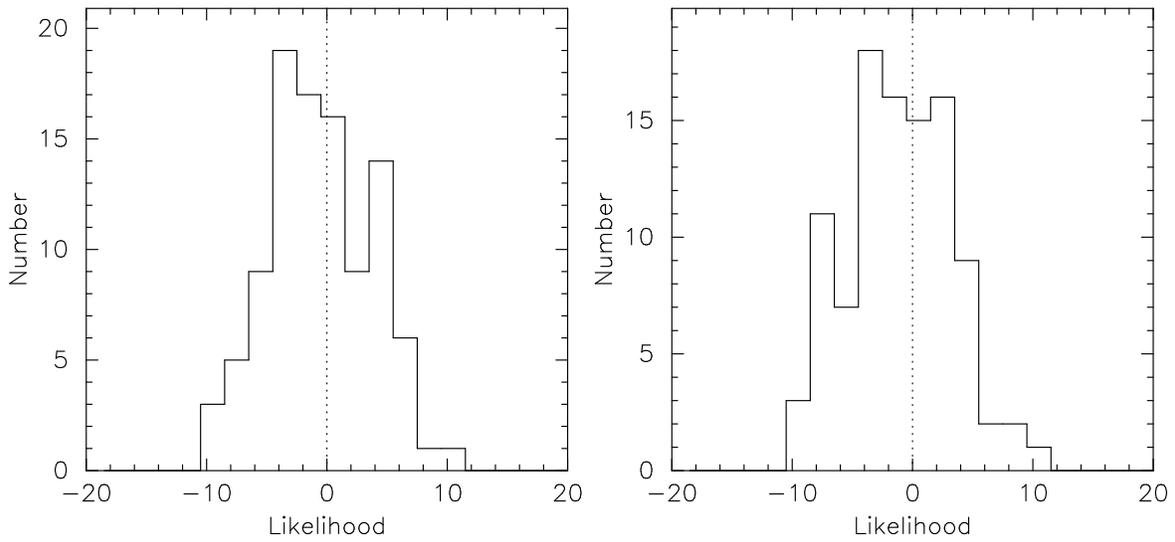

\centering
\includegraphics[scale=0.4,angle=-90]{f16a.ps}
\includegraphics[scale=0.4,angle=-90]{f16b.ps}
\caption{Histogram of the relative Bayesian maximum likelihoods calculated using 500 realizations of the best fit companion model for the symmetric Gaussian (left) and the asymmetric Gaussian (right).  The x-axis represents the maximum likelihood of each model realization compared to the best fit maximum likelihood (zero on this scale), and the y-axis gives the number of realizations at each relative likelihood.  The maximum likelihood for each realizations was compared to that derived for the best fit model.  If the model is a true representation of the data then these histograms should peak near zero, which they do.}
\label{fig:monte}
\end{figure}

\clearpage

\begin{figure}
\centering
\includegraphics[scale=0.7,angle=-90]{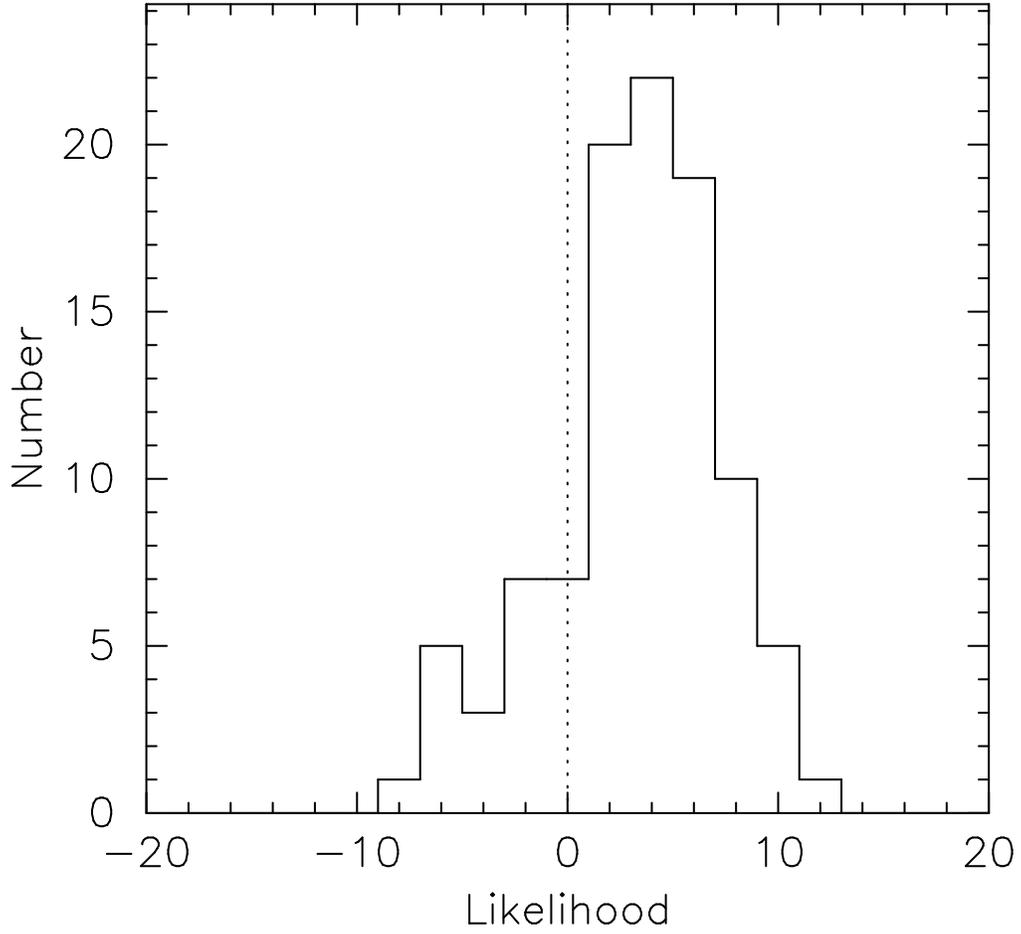}
\caption{Similar to Figure \ref{fig:monte}, but for the secondary solution to the asymmetric Gaussian model.  This demonstrates that this secondary solution is not a good fit to the data.  The likelihood of that solution (at zero on this scale) sits well off of the peak.}
\label{fig:monte2}
\end{figure}

\clearpage

\begin{deluxetable}{lccc}
\tablewidth{0pt}
\tablecaption{}
\tablehead{
\colhead{Parameter} & \colhead{Symmetric} & \colhead{Asymmetric} & \colhead{Secondary} \\
\colhead{} & \colhead{Gaussian} & \colhead{Gaussian} & \colhead{Solution}
}

\startdata
$\log(a_o)~[\log(AU)]$ & $0.86^{+0.06}_{-0.12}$ & $0.86^{+0.06}_{-0.18}$ & $0.86\pm0.06$ \\
$\sigma_a~[\log(AU)]$  & $0.28\pm0.04$ & $0.30\pm0.10$ & $0.20\pm0.02$ \\
N                      & $20\%\pm4\%$ & $22\%^{+6\%}_{-4\%}$ & $28\%\pm11\%$ \\
$\gamma$               & $1.8^{+0.4}_{-0.6}$ & $1.6\pm0.6$ & $1.4\pm1.4$ \\
\% Comp. $>$ 15~AU     & 1.1\% & 1.3\% & NA \\
\% Tight Binaries      & 2.9\% & 3.5\% & NA \\
\enddata

\end{deluxetable}

\end{document}